\documentclass[preprint,12pt, a4paper]{elsarticle}

\usepackage[utf8]{inputenc}
\usepackage[T1]{fontenc}

\usepackage{hyperref}
\usepackage{amsmath}
\usepackage{amsthm}
\usepackage{amssymb}
\usepackage{bm}
\usepackage[def-file=diffcoeff5]{diffcoeff}
\usepackage{dirtree}
\usepackage{listings}
\lstMakeShortInline[columns=fixed, basicstyle=\ttfamily]{`}

\usepackage{lineno}

\newcommand{\bz}{\bm{z}}
\newcommand{\bphi}{\bm{\phi}}
\newcommand{\btheta}{\bm{\theta}}
\newcommand{\bTheta}{\bm{\Theta}}
\newcommand{\vphi}{\varphi}
\newcommand{\bvphi}{\bm{\vphi}}
\newcommand{\bs}{\bm{s}}
\newcommand{\bh}{\bm{h}}
\newcommand{\bg}{\bm{g}}
\newcommand{\bx}{\bm{x}}
\newcommand{\by}{\bm{y}}
\newcommand{\twobyone}{2\!\times\!1}

\newcommand{\re}{\operatorname{Re}}
\newcommand{\pacpt}{p_{a}}

\newcommand{\var}[1]{\operatorname{var}\left[#1\right]}

\renewcommand{\u}[2]{{#1}^{(#2)}}
\newcommand{\avg}[1]{\left\langle{#1}\right\rangle}

\newcommand{\pytorch}{\texttt{PyTorch}}
\begin{document}
\renewcommand{\labelenumii}{\arabic{enumi}.\arabic{enumii}}

\begin{frontmatter}

\title{NeuMC - a package for neural sampling for lattice field theories}

\author[inst1]{Piotr Białas\corref{ca}}

\affiliation[inst1]{organization={Institute of Applied Computer Science, Jagiellonian University},
            addressline={ul.~Łojasiewicza~11}, postcode={30-348},
            city={Kraków},
            country={Poland}}

\author[inst2]{Piotr Korcyl}
\author[inst2]{Tomasz Stebel}

\affiliation[inst2]{organization={Institute of Theoretical Physics, Jagiellonian University},
            addressline={ul.~Łojasiewicza 11}, postcode={30-348},
            city={Kraków},
            country={Poland}}

\affiliation[sd]{organization={Doctoral School of Exact and Natural Sciences, Jagiellonian University}, addressline={ul.~Łojasiewicza 11},postcode={30-348}, city={Kraków}, country={Poland}}
            
\author[inst1,sd]{Dawid Zapolski}
\cortext[ca]{Corresponding author \texttt{piotr.bialas@uj.edu.pl}}

\begin{abstract}
We present the \texttt{NeuMC} software package, based on \pytorch, aimed at facilitating the research on neural samplers in lattice field theories. Neural samplers based on normalizing flows are becoming increasingly popular in the context of Monte-Carlo simulations as they can effectively approximate target probability distributions, possibly alleviating some shortcomings of the Markov chain Monte-Carlo methods. 
Our package provides tools to create such samplers for two-dimensional field theories. 
\end{abstract}

\begin{keyword}
Monte-Carlo \sep machine learning \sep neural networks \sep generative models
\PACS 07.05.Mh \sep 02.50.Ng
\MSC 82M31 \sep 65C05 \sep 11K45 \sep 68T07
\end{keyword}

\end{frontmatter}


\section{Introduction}

Monte-Carlo methods rely on random samples generated from some {\em target} distribution $p(\bphi)$. Often, {\em e.g.} in lattice field theories, that distribution is complicated and depends on all degrees of freedom of the system, hence there are no known methods for sampling from this distribution directly and independently. Instead, the approach of choice is to construct the associated Markov chain of samples, giving rise to the so-called Markov Chain Monte-Carlo (MCMC) approach. To be more precise, each step of the algorithm has two stages: in the first stage for the given configuration $\bphi_i$, a new trial configuration $\bphi_{trial}$ is proposed from the distribution $q(\bphi_{trial}|\bphi_i)$. In the second stage, the trial configuration is accepted with a probability $\pacpt(\bphi_{trial}|\bphi_i)$ usually given by the Metropolis-Hastings acceptance probability \cite{metropolis, Hastings}
\begin{equation}
\label{eq:pacpt}
\pacpt(\bphi_{trial}|\bphi_i)=\min\left\{1,\frac{p(\bphi_{trial})}{q(\bphi_{trial}|\bphi_i)}\frac{q(\bphi_{i}|\bphi_{trial})}{p(\bphi_{i})}\right\}
\end{equation}
In order to keep the acceptance rate high, typically the configuration $\bphi_{trial}$ differs from $\bphi_i$ only on a small subset of degrees of freedom, {\em e.g.} a single lattice site. This makes the consecutive samples highly correlated. The length of this autocorrelation, as measured in the number of Monte-Carlo samples, usually increases as we approach the critical point of the theory. This phenomenon is known as {\em critical slowing down} and is the major obstacle in obtaining reliable estimates from lattice simulations. 
Neural samplers try to alleviate this problem. The idea is to use a neural network-based model to sample data from some other distribution 
$q(\bphi|\btheta)$. The parameters $\btheta$ are then tuned as to approximate the target distribution as closely as possible. 
This distribution can subsequently be used to enhance the Monte-Carlo simulations in different  ways.

Neural samplers are a relatively new approach but have progressed quickly from the 2D Ising model \cite{VANPRL} and scalar $\phi^4$ \cite{PhysRevD.100.034515} theory, through gauge theories \cite{Kanwar2020,Abbott:2023thq} including fermions \cite{Albergo:2021bna,Albergo2022,Abbott:2022zhs}, to the 4D lattice QCD \cite{Abbott:2024kfc,abbott2024practicalapplicationsmachinelearnedflows}. Some specific applications were already proposed, {\it e.g.} the estimation of thermodynamic quantities \cite{PhysRevE.101.023304,ETOinLFT} and entanglement entropy \cite{Bialas:2024gha,Bulgarelli:2024yrz}. Also, progress has been made implementing various neural architectures; this includes diffusion models \cite{Wang:2023exq,Zhu:2025pmw}, transformers \cite{Nagai:2025rok}, stochastic normalizing flows \cite{Caselle:2022acb, Bulgarelli:2024brv} and continuous normalizing flows \cite{Gerdes:2024rjk}.

Despite rapid progress, neural samplers are still a long way from being on the level of the modern Monte-Carlo codes which have been developed over four decades, so further improvements are needed. 
Our software aims to facilitate the research in this area by providing a framework, based on a popular \pytorch~package, for building flow-based neural samplers for 2D lattice field theories. The first version of the software was successfully used while developing the REINFORCE gradient estimator and was tested on $\phi^4$ and the $U(1)$ gauge model with fermions (Schwinger model)  \cite{bialas2024, bialas2022gradientestimators}.

In this paper, we present the code which allows users to build their own samplers. The software's structure allows for easy modifications and adding new functionalities. All physical models which are currently implemented in our package were discussed in the context of neural samplers already in the literature; however, up to now to our best knowledge, there was no general framework for implementing flow-based samplers. There exists an excellent introduction to neural samplers in form of the notebook \cite{albergo2021introduction} which actually was an inspiration and the starting point for this work (some snippets of this code, taken under the CC 4.0 license, still remain in our package). A more general  framework \verb|NeuLat| was announced but it is not yet made public \cite{Nicoli:2023rcd}. 

In the next sections, we describe the general idea behind the concept of neural samples. In particular, we discuss the equivariant flows and gradient estimators and provide a description of our implementations. At the end, we provide an example using one physical model. The  \verb|NeuMC| package is hosted on GitHub \cite{repo}; instructions for installation are provided in \ref{sec:installation}.

\section{Neural samplers}

Training the distribution $q(\bphi|\btheta)$ is done by minimizing a suitable loss function that provides some measure of the difference between $q(\bphi|\btheta)$ and target distribution $p(\bphi)$.  The most common choice is the (reverse) Kullback-Leibler divergence \cite{KL}
\begin{equation}\label{eq:KL-reverse}
    D_{KL}(q|p) = \int\dl\bphi \, q(\bphi|\btheta) \left(\log q(\bphi|\btheta)-\log p(\bphi)\right).
\end{equation}
Actually, often we know the target distribution $p(\bphi)$ only up to a multiplicative constant $Z$ that we will call {\em partition function}
\begin{equation}
    P(\bphi)=Z\cdot p(\bphi).
\end{equation}
If we insert $P(\bphi)$ instead of $p(\bphi)$ in the formula \eqref{eq:KL-reverse} we obtain
\begin{equation}\label{eq:F}
    F(q|P) \equiv \int\dl\bphi \, q(\bphi|\btheta) \left(\log q(\bphi|\btheta)-\log p(\bphi)-\log Z\right)= 
    F+D_{KL}(q|p),
\end{equation}
where $F=-\log Z$ is a quantity which, with some abuse of notation, we will call {\em free energy}. As $F$ does not depend on $\btheta$, minimizing $F(q|P)$ is the same as minimizing $D_{KL}$. 

Knowledge of $Z$ is not required for training the model, but once trained the model can be used to estimate $F$, something that is very difficult to do with a traditional Monte-Carlo.  Firstly, the $F(q|P)$ provides a variational approximation of $F$ that approaches true $F$ from above because $D_{KL}$ is always non-negative. Additionally, we can also construct an unbiased estimator of $Z$ \cite{PhysRevE.101.023304, ETOinLFT}. This can be achieved using the unnormalized importance weights
\begin{equation}
\label{eq:importance-weights}
w(\bphi)\equiv\frac{P(\bphi)}{q(\bphi)} = Z\frac{p(\bphi)}{q(\bphi)}.
\end{equation} 
Obviously 
\begin{equation}
    \int\dl\bphi\, q(\bphi)w(\bphi) = Z\int\dl\bphi\, p(\bphi)=Z,
\end{equation}
so $Z$ can be approximated using 
\begin{equation}
    Z\approx\frac{1}{N}\sum_{i=1}^N w(\bphi_i),\quad \bphi_i \sim q(\bphi_i).
\end{equation}
The symbol $\sim$ is used to denote the fact that $\bphi_i$ is drawn with probability $q(\bphi_i|\btheta)$. 

\subsection{Neural Markov Chain Monte-Carlo}

Once trained, the model can be used within the Metropolized Independent Sampling (MIS) \cite{Liu} approach where the whole configuration $\bphi_{trial}$ is proposed by the model and then accepted or rejected using the Metropolis-Hastings acceptance probability
\begin{equation}
\pacpt(\bphi_{trial}|\bphi_i)=\min\left\{1,\frac{P(\bphi_{trial})}{q(\bphi_{trial}|\btheta)}\frac{q(\bphi_{i}|\btheta)}{P(\bphi_{i})}\right\}.
\end{equation}
Such an algorithm is called Neural Markov Chain Monte-Carlo (NMCMC) \cite{PhysRevE.101.023304}.
With a properly trained model, the autocorrelation can be substantially smaller than in the case of standard MCMC, {\it e.g.} Metropolis–Hastings algorithm, cluster algorithms or hybrid Monte-Carlo (see Ref.~\cite{Bialas:2021bei} for a discussion).

\subsection{Neural importance sampling}

Another approach, dubbed the Neural Importance Sampling (NIS) \cite{PhysRevE.101.023304}, uses the importance weights \eqref{eq:importance-weights}
to reweight the expectation values to the correct distribution 
\begin{equation}
    \avg{O}_p=\int\dl\bphi p(\bphi)O(\bphi)=
    \frac{\int\dl\bphi\, q(\bphi|\btheta)w(\bphi)O(\bphi)}
    {\int\dl\bphi\, q(\bphi|\btheta)w(\bphi)}.
\end{equation}
This expression can be approximated by sampling from distribution $q(\bphi|\btheta)$:
\begin{equation}
    \avg{O}_p \approx \frac{\frac{1}{N}\sum_{i=1}^N w(\bphi_i)O(\bphi_i)}
    {\frac{1}{N}\sum_1^N w(\bphi_i)},\quad \bphi_i \sim q(\bphi|\btheta).
\end{equation}

In this case all the configurations are statistically independent but the variance of weights $w$ affects the variance of the result. This can be expressed using the effective sample size (ESS):
\begin{equation}
    ESS = \frac{\avg{w}^2}{\avg{w^2}} = \frac{1}{\frac{\var{w}}{\avg{w}^2}+1},
\end{equation}
which is a measure of the quality of the sampler with regard to importance sampling  \cite{Liu, Kong} and  provides a handy, commonly used, way of measuring the progress of the training. For the perfect training the weights have zero variance and ESS is equal to one. 

\section{Physics models}
\label{sec:physics-models}

We start with a description of the lattice models we have included in our package. Formally the models are specified by providing a definition of the action $S(\bphi)$ that defines the unnormalized target distribution $P(\bphi)$
\begin{equation}
    P(\bphi) = e^{-S(\bphi)}. 
\end{equation}
Please note that this does not restrict in any way the class of target probability distributions, as every distribution can be described in this way.

Actions are defined as classes that can be instantiated with different parameters, e.g.
\begin{lstlisting}
from neumc.physics.phi4 import ScalarPhi4Action
phi4_action = ScalarPhi4Action(m2=0.5, lam=1.0) 
\end{lstlisting}

Unfortunately, defining the action is not enough.  Different theories can have different kinds of fields: scalar, vector, U(1), gauge fields, etc. requiring different treatments. This will be covered in sections \ref{sec:normalizing-flow}, \ref{sec:equivariant-flows} and \ref{sec:gauge-equivariant}. Here we introduce the types of fields that are currently handled by our package.

\subsection{\texorpdfstring{$\phi^4$}{phi4} theory - scalar and vector fields}
\label{sec:phi4-theory}

One of the simplest field theories is the $\phi^4$ theory. We provide the implementation for both the scalar and the vector theory. A $D$ dimensional vector field $\bphi$ is represented by a tensor of size 
$(N_b,D,L_x,L_y)$. The first dimension is the batch dimension, the second is the vector dimension, and the remaining ones are the spatial dimensions. Putting the vector component dimension before spatial dimensions may look strange, but this is done to conform to the \verb|PyTorch| convention for convolutional neural networks where channel dimension is before spatial dimensions. 

The $\phi^4$ action for a vector field is defined as
\begin{equation}
\begin{split}
    S(\bphi|m^2,\lambda) &= \frac{1}{2}\kappa\sum_{i,j=0}^{L-1}
    \sum_{k=0}^{D-1}\left( 
    (\phi_{k,i+1,j}-\phi_{k,i,j})^2 + (\phi_{k,i,j+1}-\phi_{k,i,j})^2\right)\\
    &\phantom{=}+\sum_{i,j=0}^{L-1}\left(\frac{m^2}{2}|\phi_{i,j}|^2+\frac{\lambda}{4!}\left(|\phi_{i,j}|^2\right)^2\right),
\end{split}    
\end{equation}
where
\begin{equation}
|\phi_{i,j}|^2 =\sum_k \phi_{k,i,j}^2.    
\end{equation}
This can be rewritten as 
\begin{equation}
\begin{split}
    S(\bphi|m^2,\lambda) &=
    -\kappa\sum_{i,j=0}^L \sum_k\left(
    \phi_{k,i+1,j}\phi_{k,i,j}
        + \phi_{k,i,j+1}\phi_{k,i,j}
    \right)\\
    &\phantom{=}+\sum_{i,j=0}^L\left(
    \frac{m^2+4\kappa}{2}|\phi_{i,j}|^2+\frac{\lambda}{4!}\left(|\phi_{i,j}|^2\right)^2
    \right)
\end{split}
\end{equation}
and this is the form we use in our implementation which is provided by the $\texttt{VectorPhi4Action}$ class. 

This action has a global $O(D)$ symmetry, where $O(D)$ is a group of orthogonal matrices of size $D$. One can easily check that transforming every element
\begin{equation}
    \phi'_{k',i,j} = \sum_{k=0}^{D-1}o_{k',k}\phi_{k,i,j},\quad \bm{o}\in O(D)
\end{equation}
does not change the action. In the scalar case this reduces to the $Z_2$ symmetry
\begin{equation}
    \phi'_{i,j} = -\phi_{i,j}. 
\end{equation}

The scalar field could be defined as a tensor of shape $(N_n,1, L_x,L_y)$ but for convenience it is "squeezed" to the shape $(N_b,L_x, L_y)$ and we provide a matching \texttt{ScalarPhi4Action} class. 

Of course, the user may provide her own implementation of the action, as long as it expects scalar or vector fields as the input. For example, one could add a regularization term. 

\subsection{XY model - periodic scalar fields}
\label{sec:XY-model}

The $XY$ model can be considered as an extension of the Ising model. The spins are generalized to two-dimensional vectors of unit length
\begin{equation}
|\bm{s}_{i,j}|=1.
\end{equation}
The action is
\begin{equation}
    -\beta\sum_{i,j=0}^{L-1} \left(\bs_{i,j}\cdot \bs_{i+1,j} + \bs_{i,j}\cdot \bs_{i,j+1}\right).
\end{equation}
Each spin can be uniquely represented by an angle
\begin{equation}
    \bs_{i,j} = (\cos\phi_{i,j}, \sin\phi_{i,j})
\end{equation}
and the action then becomes
\begin{equation}
    -\sum_{i,j=0}^{L-1} \left(\cos(\phi_{i+1,j}-\phi_{i,j})+\cos(\phi_{i,j+1}-\phi_{i,j})\right).
\end{equation}
This is the formulation we use in our implementation in \texttt{XYAction} class from module \texttt{neumc.physics.xy}.

This model has a global $O(2)$ symmetry: rotating every spin by the same angle $\varphi$
\begin{equation}
    \phi'_{i,j}=\phi_{i,j}+\varphi
\end{equation}
does not change the action.

\subsection{U(1) gauge theory}
\label{sec:u1-gauge-theory}

\begin{figure}
    \centering
    \includegraphics[width=0.35\linewidth]{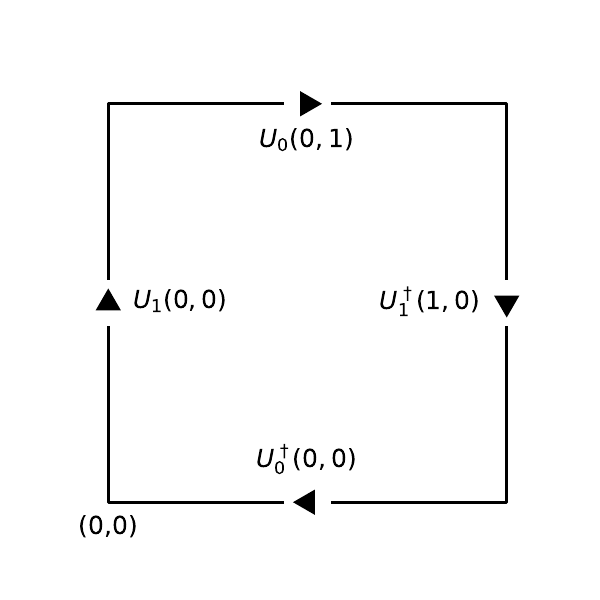}
    \caption{Scheme of plaquette.}
    \label{fig:plaquette}
\end{figure}

The most complicated model we provide is the $U(1)$ gauge theory with fermions. 
The gauge fields $U$ are defined on the {\em links} of the lattice instead of sites and are members of some gauge group $\mathcal{U}$ \cite{WilsonLatticeGauge}. By definition, we require gauge fields to respect a local gauge symmetry. The gauge symmetry is defined by a set of gauge fields $g(\bx)$ on  the sites of the lattice. If we denote by $U_\mu(\bx)$ a gauge field on a link emerging from site $x$ in the direction  $\mu$, then this link variables transform accordingly to 
\begin{equation}
\label{eq:gauge-symmetry}
    U_\mu(\bx)\rightarrow g(\bx)U_\mu(\bx)g^\dagger(\bx+\hat\mu)
\end{equation}
where $\bx+\hat\mu$ denotes a lattice site on the next site in the direction $\mu$ from $\bx$. 
The simplest action that respects this symmetry is \cite{WilsonLatticeGauge}
\begin{equation}
S(U) = -\beta\sum_{\bx}\re P(\bx)
\end{equation}
where $P(\bx)$ is a {\em plaquette} (see Figure~\ref{fig:plaquette})
\begin{equation}\label{eq:plaquette}
    P(\bx)=U_1(\bx) U_0(\bx+\hat{1}) U_1^\dagger(\bx+\hat{0}) U_0^\dagger (\bx).
\end{equation}
The versor $\hat\mu$ is the displacement vector of one lattice site in the direction $\mu$.
The gauge symmetry introduces many unphysical degrees of freedom and is usually an obstacle for training neural samplers, and we will require that the probability $q(\bphi|\btheta)$ is also invariant under its action. How to achieve this is described in Section~\ref{sec:gauge-equivariant}. 

The elements of the $U(1)$ group are just complex numbers $z$ with unit norm $|z|=1$ and can be represented by a phase angle $\phi$
\begin{equation}
    z= e^{i\phi}. 
\end{equation}
The action then becomes
\begin{equation}
    -\beta\sum_{\bx} \cos\left(\phi_1(\bx) +\phi_0(\bx+\hat{1}) -\phi_1(\bx+\hat{0}) - \phi_0(\bx)\right).
\end{equation}
and is implemented as \texttt{U1GaugeAction} class in module \texttt{neumc.physics.u1}. This module provides also many additional helper functions. 

Fermionic degrees of freedom are introduced into the theory as anticommuting Grassmann variables living on the sites of the lattice \cite{WilsonLatticeGauge}.  They cannot be represented on a computer but can formally be integrated out. After integration, we are left with the determinant of the Dirac operator. The action with two massless fermion fields thus becomes
\begin{equation}
S(U) = -\beta\sum_{\bx}\re P(\bx)-\log\det D[U]^\dagger D[U],
\end{equation}
where the  Wilson-Dirac operator $D[U]$ is defined as
\begin{equation}
\begin{split}
    D[U](\by,\bx)^{\alpha\beta}&=\delta(\by-\bx)\delta^{\alpha\beta}\\
    -\kappa & \sum_{\mu=0,1}\Big\{[1-\sigma^\mu]^{\beta\alpha}
    U_\mu(\by-\bx+\hat\mu)\delta(\by-\bx+\hat\mu)\\
    &\phantom{-\kappa\sum_{\mu=0,1}}+[1+\sigma^\mu]^{\beta\alpha}U^\dagger_\mu(\by-\hat\mu)\delta(\by-\bx-\hat\mu)
    \Big\},
\end{split}    
\end{equation}
where $\sigma^\mu$ are the Pauli matrices. The size of this matrix is $2 L^2\times 2L^2 = 4L^4$ so it grows very quickly with lattice size. At the moment we are constructing the complete matrix and are calculating the determinant directly. This unfortunately allows for only rather small lattice sizes up to $24\times 24$ \cite{bialas2024}.   This is implemented in \texttt{QEDAction} class in module \texttt{neumc.physics.schwinger}.

\section{Normalizing flow} 
\label{sec:normalizing-flow}

Normalizing flow may be defined as the tuple of functions \cite{PhysRevD.100.034515,dinh2017density,9089305}
\begin{equation}\label{eq-nf}
   \mathbb{R}^{D}\ni \bz\longrightarrow (q_{pr}(\bz),\bm{\vphi}(\bz|\btheta))\in (\mathbb{R},\mathbb{R}^{D}),
\end{equation}
where the function $q_{pr}(\bz)$ is the probability density defining a {\em prior} distribution of a random variable $\bz$. If $\bvphi(\bz|\btheta)$ is  a {\em bijection} and input $\bz$ is drawn from $q_{pr}(\bz)$ then the output $\bphi$ is distributed according to 
\begin{equation}\label{eq-q-phi}
    q(\bphi|\btheta)= q_z(\bz|\btheta) \equiv  q_{pr}(\bz)\left|J(\bz|\btheta)^{-1}\right|,\quad \bphi=\bvphi(\bz|\btheta),
\end{equation}
where
\begin{equation}\label{eq_jac_def}
    J(\bz|\btheta)=\det \left(\diffp{\bvphi(\bz|\btheta)}{\bz}\right)
\end{equation}
is the determinant of the Jacobian of $\bvphi(\bz|\btheta)$. The requirement for $\bvphi(\bz|\btheta)$ to be a bijection is needed to ascertain that $\bz$ is the only solution of the equation $\bphi=\bvphi(\bz|\btheta)$. For practical reasons, normalizing flows are constructed in such a way that the Jacobian determinant is relatively easy to compute. 

The \verb|neumc.nf| package contains utilities to build general normalizing flows as well as some  specific implementation.
In our package, the transformations $\bvphi$ are implemented as classes that derive from \verb|Transformation| abstract base class and must implement a \verb!forward! function that takes a configuration $\bm{z}$ and returns the transformed configuration $\bphi(\bz)$ together with the logarithm of the Jacobian determinant of this transformation. Coupled with a prior, this enables to sample from the distribution $q(\bphi|\btheta)$ defined by the flow as illustrated below
\begin{lstlisting}[]
    prior = ...
    transform = ... 
    z = prior.sample_n(1024)
    phi, logJ = transform(z)
    loq_q_phi = prior.log_prob(z)-logJ
\end{lstlisting}
Transformation classes may also implement an inverse transformation in the form of the \verb!reverse! function. The \verb|Transformation| class derives also from the \verb|PyTorch| \verb|torch.nn.Module| making it suitable for training. 

Very often flows are built by chaining together many transformations. To this end, we provide the \verb|TransformationSequence| class. The instances of this class are initiated with a list of transformations. The \verb|forward| and \verb|reverse| methods of this class chain together the methods from each transformation on this list. 

A feature of the normalizing flows is that we start from a prior configuration (noise) $\bz$ and then obtain both $\bphi$ and $q(\bphi|\btheta)$. Sometimes we want to obtain the probability $q(\bphi|\btheta)$ being given only the configuration $\bphi$. In this case we may use the inverse transformation $\bvphi^{-1}(\bphi)$ and use the relation 
\begin{equation}
\left|\det \frac{\partial \bvphi(\bz)}{\partial \bz}\right|^{-1}= 
\left|\det \frac{\partial \bvphi^{-1}(\bphi)}{\partial \bphi}\right|
\end{equation}
leading to
\begin{equation}
\label{eq:q-phi}
\log q(\bphi) = \log q_{pr}(\bz) +  \log \left|\det \frac{\partial \bvphi^{-1}(\bphi)}{\partial \bphi}\right|, 
\quad \bz=\bvphi^{-1}(\bphi).
\end{equation}

\subsection{Coupling layers}
\label{sec:coupling-layers}

Currently, all flows in our package are based on coupling layers (see. e.g. \cite{Kobyzev2019,Durkan2019}). The coupling layers rely on the subdivision of the lattice into active, frozen, and passive components. Only the active components are transformed while the frozen components are used to compute the parameters of the transformation, the passive components do not affect anything (the need for them will be explained in Section \ref{sec:gauge-equivariant})
\begin{equation}
\begin{split}
    \bphi'_{act.}& = \bh\left(\bphi_{act.}|
    \bTheta(\bphi_{frz.})\right)\\ 
    \bphi'_{frz.} & = \bphi_{frz.}\\
    \bphi'_{pas.}& = \bphi_{pas.}. 
\end{split}    
\end{equation}
If $\bh$, called the {\em coupling function}, is a bijection, the whole transformation is also a bijection. The function $\bTheta(\bphi_{frz.})$ is called a {\em conditioner} (we follow the notation from \cite{Kobyzev2019}) and is used to calculate the parameters of the transformation based on the frozen part of the configuration. This is actually the part where machine learning takes place, and it is usually implemented using neural networks. This function depends on the learnable parameters $\btheta$, usually the weights of the neural network, but we omit this dependence for clarity.  

The  resulting Jacobian matrix is block diagonal
\begin{equation}
    \left(\begin{array}{ccc}
           \diff{\bh\left(\bphi_{act.}|\bTheta(\bphi_{frz.})\right)}{\bphi_{act.}} &
           \diff{\bh\left(\bphi_{act.}|\bTheta(\bphi_{frz.})\right)}{\bphi_{frz.}} & 0\\
           0 & \mathbb{I} & 0\\
           0 & 0 & \mathbb{I}
    \end{array}\right). 
\end{equation}
So the Jacobian determinant is determined only by the derivatives with respect to $\bphi_{act.}$,
\begin{equation}
    J=
    \left|\det\diff{\bh\left(\bphi_{act.}|\bTheta(\bphi_{frz.})\right)}{\bphi_{act.}}\right|. 
\end{equation}
If the coupling function $\bh$ acts on $\bphi$ pointwise
\begin{equation}
    \bm{h}(\{\phi_1, \phi_2,\ldots\}|\bTheta)=\{h_1(\phi_1|\Theta_1), h_2(\phi_2|\Theta_2),\dots\}, 
\end{equation}
where $\Theta_i$ are components of the conditioner, then the Jacobian determinant is just the product of derivatives
\begin{equation}
\label{eq:J}
    J = \prod_{i=1}^N \left|\diffp{h_i(\phi_i|\Theta_i)}{\phi_i}\right|.
\end{equation}
Very often the functions $h_i$ are identical $h_i=h$ and the only difference is in the parameters $\Theta_i$ of the transformation.  

Actually, in order to avoid numerical issues in our package, instead of working with probabilities, we use their logarithms. The formula \eqref{eq:J} becomes
\begin{equation}
\label{eq:log-J}
    \log J = \sum_{i=1}^N \log \left|\diffp{h_i(\phi_i|\Theta_i)}{\phi_i}\right|.
\end{equation}

To achieve the required expressivity, many such layers are combined. The active, frozen, and passive parts are changed in each layer in some predefined fashion.  In the simplest example, there is no passive part and active and frozen parts are interchanged in each consecutive layer. 

The coupling flows are implemented in our package as \verb|CouplingLayer| class. Instances of this class are initialized with a coupling function, a conditioner, and a mask. The masks, described in more detail in the next section, are used to subdivide the configurations into active, frozen, and passive parts. The method \verb!_call! of this class is presented in the Listing~\ref{lst:coupling-layer}. It handles both forward and reverse transformations. 

This function can actually receive as an input a sequence of tensors. The first element of this sequence is considered the configuration to be transformed. The remaining elements are used only by the conditioner to calculate the parameters of the transformation. The need for this will become apparent in Section~\ref{sec:gauge-equivariant} where we describe the flows for the gauge fields. 

\begin{lstlisting}[float, label=lst:coupling-layer, caption=Implementation of the coupling layer. This function hadles both forward and reverse transformations depending on the value of the parameter \texttt{dir}.]
def _call(self, dir, xs):
    if not isinstance(xs, Sequence):
        xs = (xs,)
    x_active = xs[0] * self.mask[0]["active"]
    x_passive = xs[0] * self.mask[0]["passive"]
    
    x_frozen = [
        mask["frozen"] * item_x for mask, 
            item_x in zip(self.mask, xs, strict=True)
    ]
    parameters_ = self.conditioner(*x_frozen)
    if dir == 0:
        z_active, log_J = self.transform(
            x_active, active_mask=self.mask[0]["active"], 
            parameters=parameters_
        )
    else:
        z_active, log_J = self.transform.reverse(
            x_active, active_mask=self.mask[0]["active"], 
            parameters=parameters_
        )
    z = self.mask[0]["active"] * z_active 
                    + x_passive + x_frozen[0]
    return z, log_J
\end{lstlisting}

\subsection{Masking}
\label{sec:masking}

The subdivision into the active, frozen, and passive components is achieved using masks. The masks are defined as iterators or generators. Each call to the \verb!next! function returns a tuple containing at least one dictionary with \verb|`active`|, \verb|`frozen`|, and \verb|`passive`| keys. For each key  the corresponding value (mask) is a tensor consisting of zeros and ones. The positions of ones indicate the elements of the configuration belonging to the particular type of components (active, frozen, or passive).  Masking is done by multiplying the configuration with the mask. The coupling layers described in the previous section are constructed with masks taken sequentially from the provided iterator. Usually the masking pattern repeats after several steps (layers). For example, the checkerboard masks illustrated on the Listing~\ref{lst:masks} repeat after two steps, while some masking patterns described later repeat after four or eight steps. 
\begin{lstlisting}[float, caption=Checkerboard masks. Those two patterns are then repeated indefinitely., label=lst:masks]
masks = neumc.nf.scalar_masks.checkerboard_masks_gen((4,4))
next(masks)
({'active': tensor([[1, 0, 1, 0],
          [0, 1, 0, 1],
          [1, 0, 1, 0],
          [0, 1, 0, 1]], dtype=torch.uint8),
  'frozen': tensor([[0, 1, 0, 1],
          [1, 0, 1, 0],
          [0, 1, 0, 1],
          [1, 0, 1, 0]], dtype=torch.uint8),
  'passive': tensor([[0, 0, 0, 0],
          [0, 0, 0, 0],
          [0, 0, 0, 0],
          [0, 0, 0, 0]], dtype=torch.uint8)},)
next(mask)          
({'active': tensor([[0, 1, 0, 1],
          [1, 0, 1, 0],
          [0, 1, 0, 1],
          [1, 0, 1, 0]], dtype=torch.uint8),
  'frozen': tensor([[1, 0, 1, 0],
          [0, 1, 0, 1],
          [1, 0, 1, 0],
          [0, 1, 0, 1]], dtype=torch.uint8),
  'passive': tensor([[0, 0, 0, 0],
          [0, 0, 0, 0],
          [0, 0, 0, 0],
          [0, 0, 0, 0]], dtype=torch.uint8)},)             
\end{lstlisting}

The masks are a separate component of the coupling layer and can be varied independently. Apart from this checkerboard mask,  our package provides some more patterns that are illustrated in the \ref{sec:masking-patterns}.

\subsection{Affine coupling layers}

One of the simplest coupling functions is provided by the {\em affine coupling} \cite{Dinh2014NICENI, Dinh2016DensityEU}
\begin{equation}
    h(\phi_i|s_i, t_i) = e^{s_i}\phi_i+t_i .
\end{equation}
The logarithm of the Jacobian determinant is particularly simple in this case:
\begin{equation}
    \log J = \sum_{i} s_i .
\end{equation}

This functionality is provided by the \verb|AffineTransform| class. An example of how to construct a multilayered model is presented on the listing~\ref{lst:affine-cpl}.

\begin{lstlisting}[float, caption=Affine coupling layers. \texttt{make\_cov\_net} is an auxiliary function that constructs a convolutional neural network., label=lst:affine-cpl]
masks = neumc.nf.scalar_masks.checkerboard_masks_gen((L,L),
            device=torch_device))
layers_ = []
for i in range(n_layers):
  net = neumc.nf.nn.make_conv_net(
    in_channels=1,
    out_channels=2,
    hidden_channels=[16,16,16],
    kernel_size=3,
    dilation=1
  )
  conditioner = neumc.nf.affine_cpl.
    AffineScalarNetConditioner(net)
  cpl_function = neumc.nf.affine_cpl.AffineTransform
  
  layer =  neumc.nf.coupling_flow.CouplingLayer(
            conditioner=conditioner,
            transform=cs_coupling,flow
            mask=next(masks) )
  layers_.append(layer)
layers = neumc.nf.flow_abc.FlowSequence(layers_)
layers.to(device=torch_device);

\end{lstlisting}

\subsection{Flows on a circle}

In our package, we implement several physical models. Among them are the $XY$ and $U(1)$ gauge models. In both of these cases, the field variables can be represented by phase $\vphi\in [0,2\pi)$, so the coupling function $h$ should be a diffeomorphism of a unit circle into itself \cite{Rezende2020}. For $h$ to be a diffeomorphism it is sufficient that
\begin{align}
    h(0)&=0,\label{eq:diff1}\\
    h(2\pi)&=2\pi,\\
    \diff{h(\vphi)}{\vphi}&>0,\\
    \diff{h(\vphi)}{\vphi}[\vphi=0]&= \diff{h(\vphi)}{\vphi}[\vphi=2\pi]\label{eq:diff4}
\end{align}

If we have a set of such transformations $h_i(\vphi)$ satisfying conditions \eqref{eq:diff1}-\eqref{eq:diff4} then the convex combination  of those transformations 
\begin{equation}\label{eq:convex}
    h(\vphi)=\sum_i\rho_i h(\vphi),\qquad \rho_i\geq 0,\;\sum_i \rho_i =1 
\end{equation}
also satisfies those conditions. In this way, we can form more complex transformations from simpler ones.

The phase variables used to represent the fields are "discontinuous" in that sense that they do not convey the periodicity of the fields. Values zero and $2\pi$  are distinct but they describe the same angle. That may hinder the training of the neural networks. Because of that, as the input to the neural network we will use the representation in terms of a vector
\begin{equation}\label{eq:u1-vector}
(\cos \phi, \sin \phi)
\end{equation}
that also uniquely determines the phase $\phi$ but is periodic and continuous in $\phi$. 

In our package we implement two transformations: non-compact projection and circular splines \cite{Rezende2020}. 

\subsubsection{Non-compact projection}

\begin{figure}
    \centering
    \includegraphics[width=0.65\linewidth]{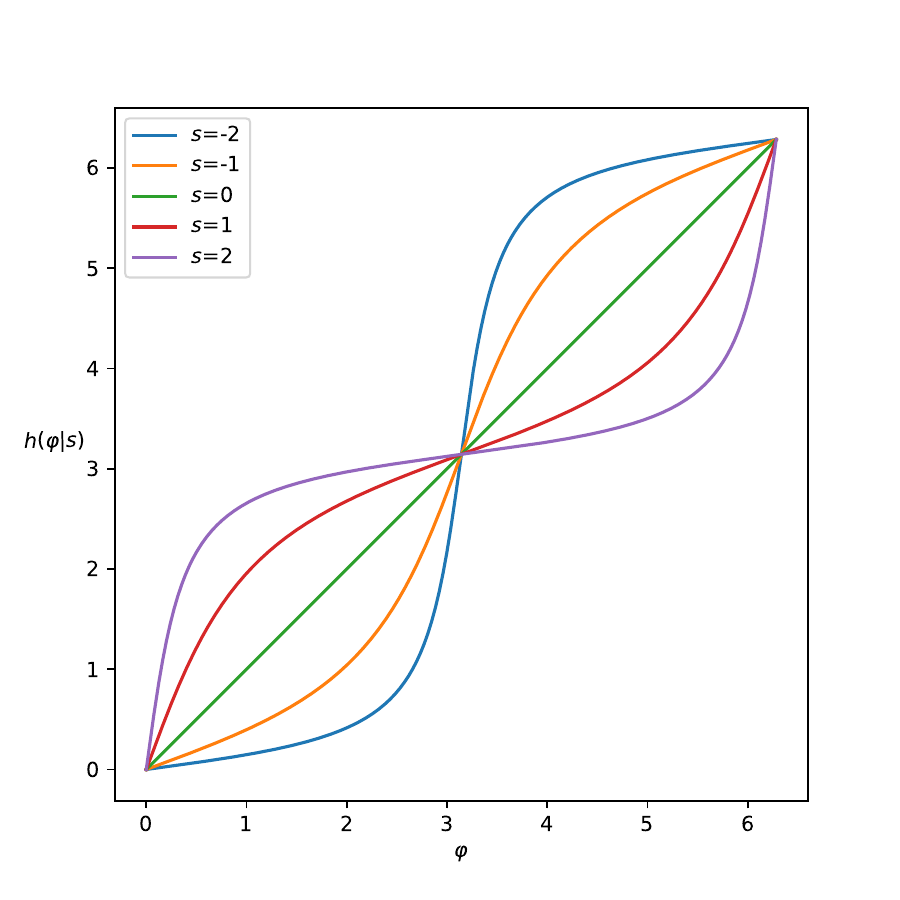}
    \caption{The non-compact projection transformation for various values of $s$.}
    \label{fig:ncp}
\end{figure}

The idea of non-compact projection is to first project the compact variable $\vphi\in(0,2\pi)$ onto $\mathbb{R}$ using transformation 
\begin{equation}
\label{eq:ncp}
x(\vphi)=\tan\left(\frac{\vphi}{2}\right)
\end{equation}
and next transform $x$ by a linear transformation 
\begin{equation}
\label{eq:ncp-linear}
    g(x)=e^s x
\end{equation}
finally projecting it back onto $(0,2\pi)$ using transformation $x^{-1}(\vphi)$. This leads to final expression
\begin{equation}
    h(\vphi|s)= 2 \tan^{-1}\left(
    e^s \tan\left(\frac{\vphi}{2}\right)\right)
\end{equation}
with derivative
\begin{equation}
    \diff{h(\vphi|s)}{\vphi}=
    \left[
    e^{-s}\sin^2\left(\frac{\vphi}{2}\right)+e^s\cos^2\left(\frac{\vphi}{2}\right)
    \right]^{-1}
\end{equation}
An example of such transformations is presented in Figure~\ref{fig:ncp}. 

Using the convex combination \eqref{eq:convex} we can combine several of those transforms together also adding a final shift $t$
\begin{equation}
    h(\vphi|\bm{s})=\mod\left(\frac{1}{N}\sum_{i=1}^N h(\vphi|s_i) + t,2\pi\right).
\end{equation}
This transformation is guaranteed to be a bijection, however, there is no closed formula for the inverse transform. The inverse can be calculated numerically using the binary search, as the function is strictly monotonic. We do not currently implement the \texttt{reverse} method for this transformation. 

This transformation differs slightly from the formulation in \cite{Rezende2020}. Firstly, we use a linear transformation \eqref{eq:ncp-linear} instead of an affine one. This slight reduction of expressivity is compensated by combining many such transformations. Secondly, we do not subtract $\pi/2$ from the argument of the projection \eqref{eq:ncp}. This only shifts the origin and does not matter as we add a random shift at the end and wrap the result to the interval $[0,2\pi)$.  

\subsubsection{Circular splines}

\begin{figure}
    \centering
    \includegraphics[width=0.66\linewidth]{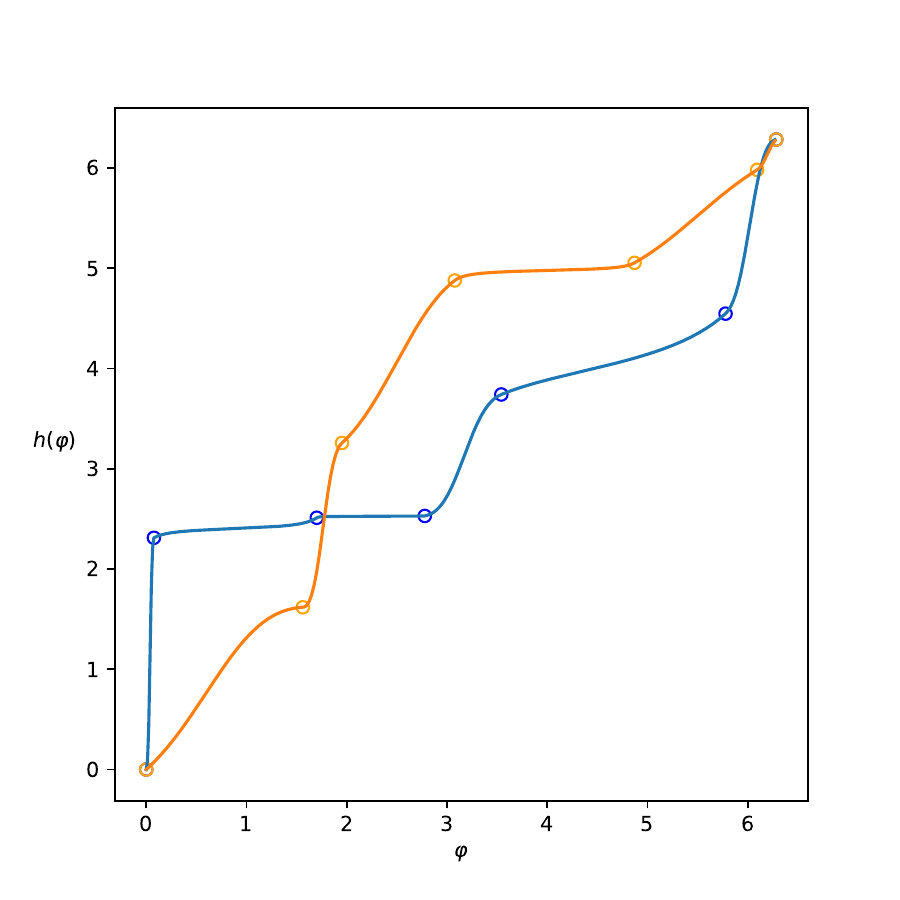}
    \caption{Circular splines with six knots.}
    \label{fig:circular-splines}
\end{figure}

Circular splines are a special example of {\em monotonic rational-quadratic splines} \cite{Durkan2019}.  A spline is defined by a set of monotonically increasing {\em knots} $\{(\u{x}k,\u{y}k)\}_0^K$ together with a positive derivative $\u\delta{k}>0$ at each knot.  Circular splines require that 
\begin{equation}
    \u{x}0=\u{y}0=0,\quad \u{x}K=\u{y}K=2\pi,\quad\u\delta{0}=\u\delta{K}.
\end{equation}
Two examples of circular splines with seven knots is presented in Figure~\ref{fig:circular-splines}.
Within our package the splines are parameterized by three sets of $K$ positive numbers: widths $\{\u{w}k\}_{k=1}^K$, 
heights $\{\u{h}{k}\}_{k=1}^K$ and derivatives $\{\u{s}{k}\}_{k=1}^K$. Widths and heights are such that
\begin{equation}
    \sum_{k=1}^K \u{w}k = \sum_{k=1}^K \u{h}k = 1.
\end{equation}
Then, we set 
\begin{align}
    \u{x}0&=0,\; \u{x}k = \u{x}{k-1}+2 \pi \u{w}k\\
    \u{y}0&=0,\; \u{y}k = \u{y}{k-1}+2 \pi \u{h}k\\
    \u\delta0&=\u{s}K, \u\delta{k} = \u{s}k
\end{align}
for $k=1,2,\ldots,K$. 
The actual formulas of the transformation $h$, its inverse, and the derivatives are much more involved than in the case of NPC and can be found in \cite{Durkan2019}. We have implemented them in the module \verb|neumc.splines.cs|. We provide some details of this implementation in \ref{sec:rational-splines}.

\subsection{Equivariant flows}
\label{sec:equivariant-flows}

Often, if not always, the physical models possess some symmetry. Those symmetries can be represented as a group $\mathcal{G}$ with elements acting on the configuration $\phi$
\begin{equation}
    \phi'= g(\phi),\qquad g\in \mathcal{G}.
\end{equation}
The group $\mathcal{G}$ is a symmetry of the model if the target probability distribution is invariant under the action of its elements
\begin{equation}
p(g(\bphi))=p(\bphi)\qquad g\in\mathcal{G}.
\end{equation}
We will also require that 
\begin{equation}
\label{eq:unitary}
    \left|\det \diffp{\bg(\bphi)}{\bphi}\right|=1
\end{equation}
which is automatically satisfied if $\bg$ is a unitary transformation. 

It is advantageous for the distribution $q(\bphi)$ to also respect this symmetry. In principle, the model could learn the symmetry by itself during the training. In practice, however, it may be unable to sample all the symmetry sectors leading to broken symmetries and mode collapse. Also when the symmetry group  is very big, this puts an additional burden on the model. 

One way to construct the flows that respect the symmetry is by the use of equivariant flows. An equivariant transformation $\bh$ commutes with the symmetry group
\begin{equation}
    \bh(\bg(\bphi_{act.})|\bTheta(\bg(\bphi_{frz.})))= \bg(\bh(\bphi_{act.}|\bTheta(\bphi_{frz.}))),\qquad \bg\in\mathcal{G}.
\end{equation}
If we pair this transformation with an invariant prior
\begin{equation}
    q_{pr}(\bg(\bz))=q_{pr}(\bz)
\end{equation}
we obtain an invariant distribution $q(\bphi)$.

The simplest example of such equivariance is the $\mathbb{Z}_2$ symmetry: $\bg(\bphi)=-\bphi$. In this case, the equivariance requires the $\bh$ to be an odd function: $\bh(-\bphi)=-\bphi$. 

\subsection{Gauge equivariant layers}
\label{sec:gauge-equivariant}

A very important class of symmetries are the gauge symmetries \eqref{eq:gauge-symmetry} described in Section~\ref{sec:u1-gauge-theory}. 

We provide an equivariant gauge layer as defined in \cite{Kanwar2020}. The idea is that when we want to transform a link $U_\mu(\bx)$ we first form an open (untraced) loop
\begin{equation}
    L(\bx) = U_{\mu}(\bx)S(\bx+\hat\mu,\bx)
\end{equation}
where $S(\bx,\bm{y})$ is some path that starts at $\bx$ and ends in $\bm{y}$.  Except for the case of abelian gauge groups like $U(1)$, this is not an invariant and under gauge symmetry this loop transforms 
as 
\begin{equation}
    L(\bx)\rightarrow g(\bx)L(\bx)g^\dagger(\bx)
\end{equation}
so we need to find a transformation, called a kernel in \cite{Kanwar2020},  that commutes with this symmetry
\begin{equation}
    h(g(\bx)L(\bx)g^\dagger(\bx)) = g(\bx)h(L(\bx))g^\dagger(\bx). 
\end{equation}
At the moment, we provide only the implementation for the gauge group $U(1)$. As this is an abelian group, the relation above is trivially satisfied. For the general case of $SU(N)$, see \cite{Boyda2021}.  

After transforming the loop, the link  $U_{\mu}(\bx)$ is modified in such a way that the loop value changes to $h(L(\bx))$. This can be done by setting
\begin{equation}
\label{eq:active-link}
U'_\mu(\bx) = h(L(\bx))S^\dagger(\bx+\hat{\mu},\bx). 
\end{equation}
Because
\begin{equation}
    S(\bx+\hat\mu,\bx)=U_\mu^\dagger(\bx) L(\bx) 
\end{equation}
we can rewrite \eqref{eq:active-link} as
\begin{equation}
\label{eq:active-link-update}
U'_\mu(\bx) = h(L(\bx))L^\dagger(\bx)U_\mu(\bx).
\end{equation}

The general loop transformation $\bh(\bm{L}|\bTheta)$ is implemented as a coupling layer. As in other coupling layers,  the kernel $\bh(\bm{L}|\bTheta)$ will transform only a subset of loops and use the rest to calculate the parameters of the transformation $\bTheta$. We will demand that $\bTheta$ depend only on the quantities that are invariant under the gauge transform \eqref{eq:gauge-symmetry}
\begin{equation}
\bTheta(\bphi_{frz.}) =  \bTheta(\bm{I}(\bphi_{frz.}))  
\end{equation}
Here $\bm{I}$ stands for any set of gauge invariant quantities constructed from fields $\bphi_{frz.}$. In general, they will be traces of some loops.

The subdivision of the configuration into active and frozen parts is slightly more involved than simple masks described previously. 
First of all, we will have a set of active links that are to be transformed by the layer. Corresponding to that, we will have a subset of loops that will be used to change those links according to \eqref{eq:active-link-update}. We will call them {\em active} loops. 
Those loops cannot contain other active links. Then there will be a set of loops that do contain the active links but were not used to transform them. We will call those loops {\em passive}, as they will not play any role in the transformation. And finally, there will be loops that do not contain any of the active links. Those loops will be called {\em frozen} and will be used to form gauge invariants $\bm{I}(\bphi_{frz.})$ used to calculate the parameters $\bTheta$ of the kernel $\bh$.

The exact way of subdividing the lattice will depend crucially on the type of loops we use. The simplest loop is a plaquette \eqref{eq:plaquette}. 
\begin{figure}
    \centering
    \includegraphics[width=\linewidth]{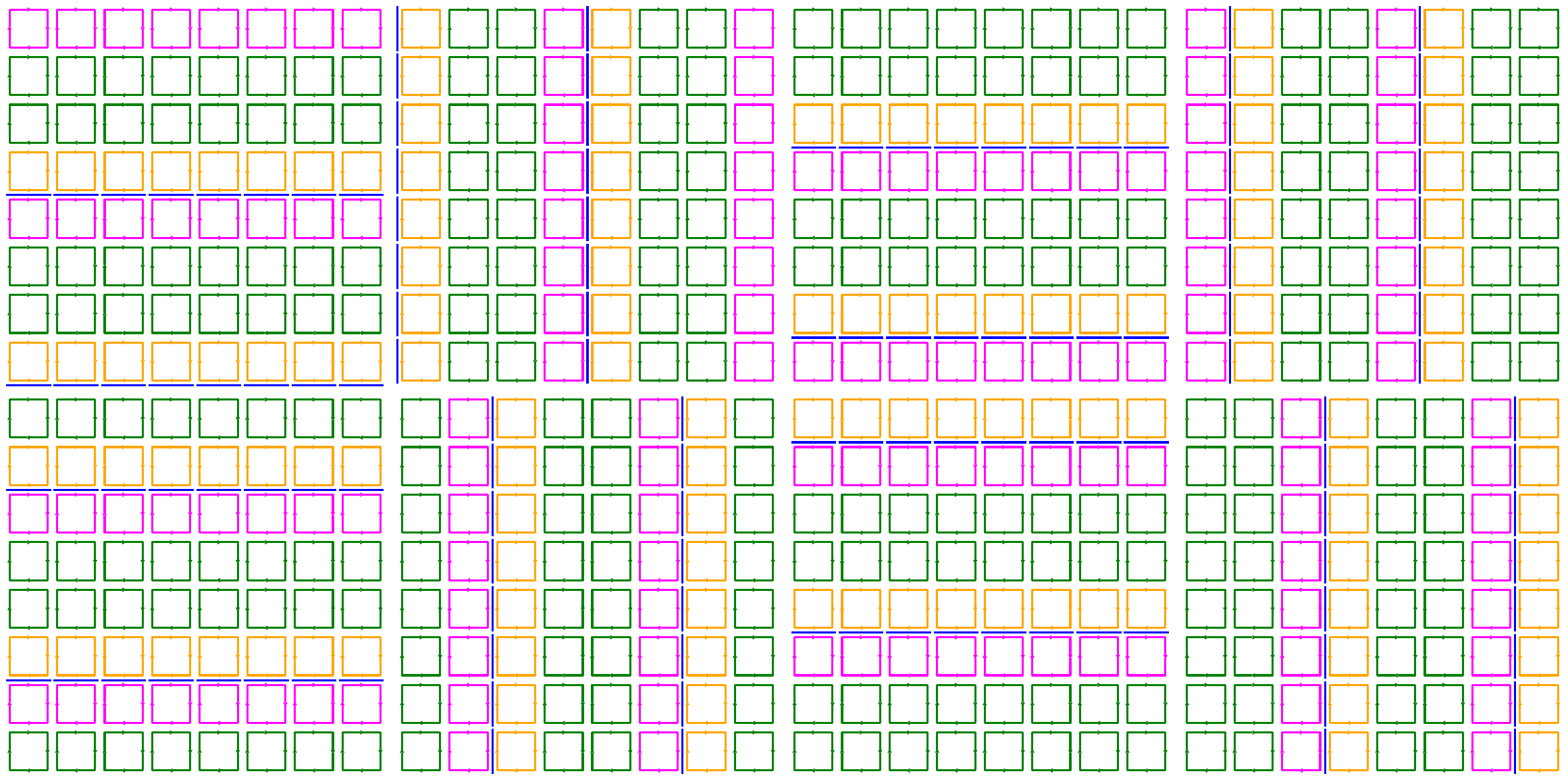}
    \caption{Simple $U(1)$ masks. Active plaquettes are marked in orange, frozen sites are in green, and passive in magenta. 
    Active links are marked in blue. 
    The period is eight.}
    \label{fig:masks-u1}
\end{figure}
A masking pattern based on plaquettes, as proposed in \cite{Kanwar2020}, is presented in Figure~\ref{fig:masks-u1}. Another masking pattern used in \cite{Albergo2022} is presented in \ref{sec:gauge-fields-masks}.

The $U(1)$ gauge equivariant coupling layer is implemented as \texttt{U1GaugeEquiv\-CouplingLayer} class. This class is responsible for updating the active links. To this end, it calculates the necessary loops that are subsequently passed to a coupling layer for transformation. After that, the class handles the update of the active links using the formula \eqref{eq:active-link-update} (see Listing~\ref{lst:u1-gauge-equiv}).   

\begin{lstlisting}[float, caption=U1 gauge equivariant layer., label=lst:u1-gauge-equiv]
def forward(self, z):
    assert torch.all(z < 2 * torch.pi)
   
    plaq = compute_u1_plaq(z, mu=0, nu=1)
    
    if self.loops_function:
        loops = self.loops_function(z)
    else:
        loops = []

    new_plaq, logJ = self.plaq_coupling([plaq] + loops)
    
    delta_plaq = new_plaq - plaq
    delta_links = torch.stack(
        (delta_plaq, -delta_plaq), dim=1
    ) 
    fx = (
        self.active_links_mask * torch_mod(delta_links + z)
        + (1 - self.active_links_mask) * z
    )

    return fx, logJ    
\end{lstlisting}

In general, loops do not have to be restricted to plaquettes. While this is a reasonable choice for active loops, it may be beneficial to use a larger set of invariants when calculating the parameters of the transformation. To allow for that, the \verb|_call| method of the generic coupling layer class described in \ref{sec:coupling-layers} can take a sequence of tensors as an input. Only the first tensor in the sequence is used to calculate the active part. All others are used only as frozen parts. Consequently, we must also provide a frozen mask for each of those additional tensors. We give an example of such additional loops in \ref{sec:masking-patterns}. 

\section{Stochastic gradient descent}

As described at the beginning, we tune the flow parameters $\btheta$ by minimizing some loss function. This is done
 using some form of the stochastic gradient descent algorithm, so at each step an estimate of the gradient of the loss function with respect to the model parameters is required. While the heavy lifting is done using the autodifferentiation capabilities of the \verb|PyTorch| library, there are different ways of constructing such a gradient {\em estimator}, each having different impact on the training \cite{bialas2024, bialas2022gradientestimators, pathGradient}.   Our package abstracts this in the form of \texttt{GradientEstimator} class.  This class is initialized with a prior, transformation, and action. 
Each subclass of \verb!GradientEstimator! class must provide a \verb!forward_pass! function. This function takes a configuration $\bz$ and the logarithm of its probability. It applies the transformation to $\bz$  to obtain $\bphi$ and returns a "loss", a logarithm of $q(\bphi|\btheta)$ and  a logarithm of $p(\bphi)$. The loss is constructed in such a way that running \verb|backward| on it will calculate the required gradients with respect to $\btheta$. 

\subsection{Reparameterization trick}
The simplest estimator is based on the {\em reparameterization trick}. It consists of rewriting the loss function \eqref{eq:F} as an integral over $\bz$ rather than over $\bphi$
\begin{equation}\label{eq:F-rt}
    F(q|P) = \int\dl\bz\, q_{pr}(\bz)\, q(\bvphi(\bz)|\btheta) \left(\log q(\bvphi(\bz)|\btheta)-\log P(\bvphi(\bz))\right).
\end{equation}
This can be approximated by the average
\begin{equation}
    F(q|P)\approx \frac{1}{N}\sum_{i=1}^N\left(\log q(\bvphi(\bz_i)|\btheta)-\log P(\bvphi(\bz_i))\right),\quad \bz_i\sim q_{pr}(\bz_i)
\end{equation}
Because $q_{pr}(\bz)$ does not depend on $\btheta$, this expression can directly be used as the loss function and its gradient can be automatically calculated using the autodifferentiation capabilities of the \verb|PyTorch| package. The implementation of this estimator is presented in the Listing~\ref{lst:rt}.
\begin{lstlisting}[float,caption=Reparameterisation trick gradient estimator., label=lst:rt]
def forward_pass(self, z, log_prob_z):
    x, log_J = self.flow(z)
    logq = log_prob_z - log_J

    logp = -self.action(x)
    loss = dkl(logp, logq)

    return loss, logq.detach(), logp.detach()
\end{lstlisting}


\subsection{REINFORCE}

REINFORCE algorithm \cite{ bialas2022gradientestimators} relies on differentiating the formula \eqref{eq:F} directly without any reparameterization  and the final formula is \cite{VANPRL, Rezende2016}
\begin{equation}\label{eq:def-g2}
   \frac{1}{N}\sum_{i=1}^N 
   \diffp{ \log q(\bphi_i|\btheta)}{\btheta}  \left(s(\bphi_i|\btheta)-\overline{s(\bphi|\btheta)_N} \right),
\end{equation}
where 
\begin{equation}\label{eq:signal}
     s(\bphi|\btheta) \equiv  \log q(\bphi|\btheta)-\log P(\bphi)\quad\text{and}\quad  \overline{s(\bphi|\btheta)_N} =\frac{1}{N}\sum_{i=1}^{N} s(\bphi_i|\btheta).
\end{equation}
This formula requires the knowledge  of $q(\bphi|\btheta)$ that we can obtain indirectly by reversing the flow and using the  formula~\eqref{eq:q-phi}. It is important that only $q(\bphi|\btheta)$ is differentiated, so we have to switch off the gradient calculations while evaluating the signal $s(\bphi|\btheta)_N$. The resulting implementation is presented in the Listing~\ref{lst:reinforce} \cite{bialas2024, bialas2022gradientestimators}. The advantage of this estimator is that it does not require the gradients to propagate through the action calculation. That provided a significant speed-up  in case of the Schwinger model \cite{bialas2024}.

\begin{lstlisting}[float, caption=REINFORCE gradient estimator., label=lst:reinforce]
def forward_pass(self, z, log_prob_z):
    with torch.no_grad():
        phi, log_J = self.flow(z)
        logq = log_prob_z - log_J
        logp = -self.action(phi)
        signal = logq - logp

    z, log_J_rev = self.flow.reverse(phi)
    prob_z = self.prior.log_prob(z)
    log_q_phi = prob_z + log_J_rev
    loss = torch.mean(log_q_phi * 
                    (signal - signal.mean()))

    return loss, logq, logp    
\end{lstlisting}

\subsection{Path gradient}
\newcommand{\bn}{\blacktriangledown}
Path gradient formulation is more involved, the final formula being \cite{pathGradient}
\begin{equation}
\frac{1}{N}\sum_{i=1}^N\left(\bn_\theta S(\bvphi(\bz_i|\btheta))+\bn_\theta\log q(\bvphi(\bz_i|\btheta)|\btheta)\right),\quad \bz_i\sim q_{pr}(\bz_i),
\end{equation}
where the path gradient $\bn_\theta$ is defined as
\begin{equation}
    \bn_\theta f(\bvphi(\bz)|\btheta)\equiv \diffp{f(\bvphi(\bz|\btheta)|\btheta)}{\bvphi(\bz|\btheta)}\diffp{\bvphi(\bz|\btheta)}{\btheta}.
\end{equation}
Because the action $S$ does not depend explicitly on $\btheta$, its path gradient is equal to a normal derivative
\begin{equation}
    \bn_\theta S(\bvphi(\bz_i|\btheta)) = \diff{S(\bvphi(\bz_i|\btheta))}{\btheta}.
\end{equation}
So the final formula is
\begin{equation}
\label{eq:path-gradient}
\frac{1}{N}\sum_{i=1}^N\left(\diff{S(\bvphi(\bz_i|\btheta))}{\btheta}
 +
\underbrace{\diffp{\log q(\bvphi(\bz_i|\btheta)|\btheta)}
{\bvphi(\bz_i|\btheta)}}_{G_i}
\diffp{\bvphi(\bz_i|\btheta)}{\btheta}
\right).
\end{equation}
Our implementation is presented in Listing~\ref{lst:path-gradient}. 
The second term in \eqref{eq:path-gradient} requires a contraction of the gradient $G_i$ with respect to fields $\bphi$ with the gradient of $\bphi$ with respect to the parameters $\btheta$. To calculate $G_i$ we first switch off the automatic gradient calculations of the derivatives with respect to $\btheta$ (line 2) and set up calculations of gradients with respect to $\bphi$ (line 5). We then calculate those gradients (lines 6-10). After switching on (line 12) the automatic differentiation with respect to $\btheta$, we contract $G$ with $\bphi$ (line 16). The subsequent call to \verb|backward| will calculate the required gradient.  

\begin{lstlisting}[float, caption=Path gradient estimator. We calculate the gradient $G_i$ on lines 2-10. We first switch of the automatic differentiation with respect to $\bphi$ on line 2 and set up the calculation of gradients with respect to $\bphi$ on line 5. On line 12 we restore the calculation of derivatives with respect to $\btheta$ and contract $G_i$ with $\bphi$ on line 16. The subsequent call to \texttt{backward} will calculate the required gradient.  
,label=lst:path-gradient, numbers=left]
def forward_pass(self, z, log_prob_z):
    nf.detach(self.flow)
    with torch.no_grad():
        fi, _ = self.flow(z)
    fi.requires_grad_(True)
    zp, log_J_rev = self.flow.reverse(fi)
    prob_zp = self.prior.log_prob(zp)
    log_q = prob_zp + log_J_rev
    log_q.backward(torch.ones_like(log_J_rev))
    G = fi.grad.data
    
    nf.attach(self.flow)
    fi2, _ = self.flow(z)
    log_p = -self.action(fi2)
    axes = tuple(range(1, len(G.shape)))
    contr = torch.sum(fi2 * G, dim=axes)
    loss = torch.mean(contr - log_p)
    return loss, log_q.detach(), log_p.detach()   
\end{lstlisting}


\subsection{Reverse training}
Given the prior, flow and gradient step, a basic training loop is illustrated in Listing~\ref{lst:training-loop}.
\begin{lstlisting}[float, label=lst:training-loop, caption=A rudimentary training loop.]
    estimator = PathGradientEstimator(
                    prior, transform, action)
    for e in range(n_epochs):
        optimizer.zero_grad()
        z = prior.sample_n(batch_size=1024)
        log_prob_z = prior.log_prob(z)
        loss, log_q, log_p = 
                estimator.forward_pass(z, log_prob_z)
        loss.backward()
        optimizer.step()
\end{lstlisting}

The \verb|GradientEstimator| class implements also an auxiliary method \verb|step|. This method encapsulates the operations presented in the Listing~\ref{lst:training-loop} making it easier to use. It can also generate several batches running \verb|forward_pass| more than once. Because it does not call the \verb|zero_grad| method of the optimizer, between those calls the gradients are accumulated. In this way, we can effectively use larger batches even if the calculations do not fit on a single GPU. This is illustrated in the Listing~\ref{lst:step}. The result of running this code would be the same as the code from the Listing~\ref{lst:training-loop},  but the calculations would be split into four batches of 256 configurations. 
\begin{lstlisting}[float, label=lst:step, caption=Using the \texttt{step} method of GradientEstimator.]
    estimator = PathGradientEstimator(
                    prior, transform, action)
    for e in range(n_epochs):
        optimizer.zero_grad()
        loss, log_q, log_p = 
                estimator.step(batch_size=256, n_batches=4)
        optimizer.step()
\end{lstlisting}

\subsection{Forward training}
\label{sec:forward-training}

So far we have been training the models in the {\em self-supervised} way, that is the model was used to generate the data that was later used in training. To train the model  a reverse  Kullback-Leibler divergence \eqref{eq:KL-reverse} was used, as we were generating data from  distribution $q(\bphi|\btheta)$.
Sometimes it may be advantageous to train the model on the data sampled from the target distribution $p(\bphi)$ \cite{Hackett:2021idh}. To this end the  {\em forward} KL divergence can be used. The forward KL divergence is defined as
\begin{equation}
\label{eq:KL-forward}
D_{KL}(p|q) = \int\text{d}\bphi\,p(\bphi)\left(\log p(\bphi) -\log q(\bphi|\btheta)\right).
\end{equation}
and differs from \eqref{eq:KL-reverse} by the order of the arguments. 
Because distribution $p(\bphi)$ does not depend on parameters $\btheta$, minimizing $D_{KL}(p|q)$ amounts to minimizing 
\begin{equation}
\label{eq:KL-loss}
-\int\text{d}\bphi\,p(\bphi)\log q(\bphi|\btheta) \approx  -\frac{1}{N}\sum_{i=1}^N \log q(\bphi^i|\btheta),\quad \bphi^i\sim p(\bphi^i).
\end{equation}
This requires a reversible flow so that $q(\bphi|\btheta)$ can be calculated according to \eqref{eq:q-phi}.

Some rudimentary functionality to facilitate this kind of training is provided in the 
\texttt{ForwardGradientEstimator} class in module \texttt{neumc.training\-.forward}. As in the case of other gradient estimators, this class provides the \verb|forward_pass| method. The input to this method is a batch of configurations generated from the distribution $p(\bphi)$. The user is responsible for feeding the batched configurations to the estimator, which we do here using a dataloader provided in  \verb|PyTorch|.  This is illustrated in the Listing~\ref{lst:forward}. 
\begin{lstlisting}[float, caption=Forward training., label=lst:forward]
from neumc.training.forward import \
        ForwardGradientEstimator

dataset = torch.utils.data.TensorDataset(cfgs)
dataloader = torch.utils.data.DataLoader(
        dataset=dataset, shuffle=True, 
        batch_size=batch_size, drop_last=True)

optimizer = torch.optim.Adam(layers.parameters(), 
                             lr=0.001)
grad_estimatior = ForwardGradientEstimator(
    prior=prior, flow=layers, 
    action=phi4_action, device=torch_device)

for epoch in range(4):
  print(epoch)
  for (phi,) in dataloader:
    optimizer.zero_grad()
    phi = phi.to(torch_device)
    loss, log_q, log_p = grad_estimator.forward_pass(phi)
    loss.backward()
    optimizer.step()
\end{lstlisting}

\section{Example}
\label{sec:examples}

We have already provided some code examples throughout the paper. In the listings below, we present a complete "bare-bones" implementation of the training script for the Schwinger model. The code is taken from \verb|schwinger.py| script in the \verb|scripts|  directory. 

We start by initializing some base parameters
\begin{lstlisting}[label=lst:init]
import torch
import neumc
torch_device = 'cuda'
float_dtype = torch.float32

L = 8
lattice_shape = (L, L)

qed_action = neumc.physics.schwinger.QEDAction(
    beta=1.0, 
    kappa=0.276)
\end{lstlisting}

\begin{figure}
    \centering
    \includegraphics[width=0.35\linewidth]{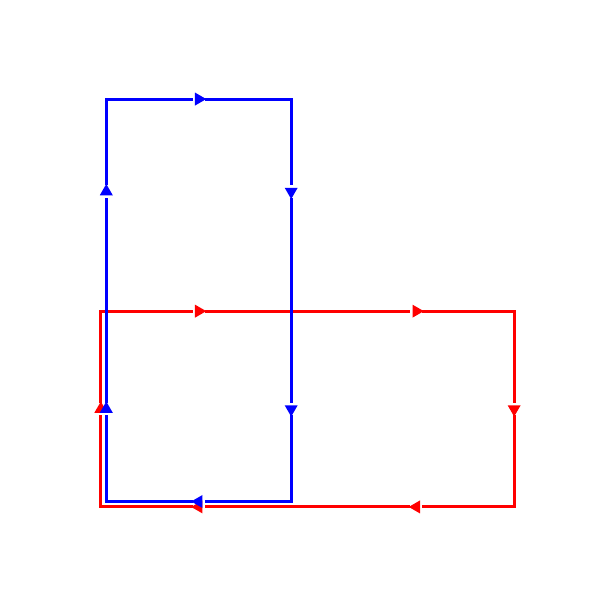}
    \caption{$2\times 1$ loops.}
    \label{fig:2x1-loops}
\end{figure}

Next, we define the masking patterns that will be responsible for dividing the configurations into active, frozen, and passive parts. We choose the masking pattern as defined in \cite{Albergo2022} and presented in \ref{sec:gauge-fields-masks} in Figure~\ref{fig:masks-sch}
\begin{lstlisting}[]
masks = neumc.nf.gauge_masks.sch_2x1_masks_gen(
    lattice_shape=lattice_shape,
    float_dtype=float_dtype,
    device=torch_device,
)
loops_function = lambda x: \
            [neumc.physics.u1.compute_u1_2x1_loops(x)]
in_channels = 6
\end{lstlisting}
We define the loops function that will calculate the additional $\twobyone$ loops (see Figure~\ref{fig:2x1-loops}) that will be subsequently used as additional input to the conditioner. This choice determines also the number of input channels of the neural network used by the conditioner. We have three types of loops: plaquettes and two orientations of the $\twobyone$ loops. Each loop is a $U(1)$ variable that is represented by a phase $\phi$, but  we use sine/cosine representation \eqref{eq:u1-vector} as the input to the neural network which brings the total number of channels to six. 

Next, we define the flow. We will use the circular splines with nine knots to implement it. We set the necessary parameters.
\begin{lstlisting}[]
n_knots = 9
out_channels = 3 * (n_knots - 1) + 1
n_layers = 48
\end{lstlisting}
The number of knots determines the number of parameters of the transformation, and hence the number of output channels of the neural network used in the conditioner. We set the number of layers to 48. When paired with the masks, this implies that every link is updated six times with one pass through the flow. 

And finally we have to set the parameters of the neural network inside the conditioner.
\begin{lstlisting}
hidden_channels = [64, 64]
kernel_size = 3
dilation = [1, 2, 3]
\end{lstlisting}
The \verb|hidden_channels| variables indicated the number of channels of the hidden layers within the neural network, so our network will have three layers in total: two hidden and one output layer. The next parameter specifies the kernel size. If it is an integer, the same kernel size will be used for all the layers. If we provide a list, we can specify different kernel sizes for each layer. The \verb|dilation| parameter works the same way, here we specify different dilation for each layer of the network \cite{Albergo2022}. 

This done, we may finally construct the needed transformation.
\begin{lstlisting}[]
import neumc.nf.cs_coupling as csc
from neumc.nf.u1_equiv import U1GaugeEquivCouplingLayer

layers = []
for l in range(n_layers):
    net = neumc.nf.nn.make_conv_net(
        in_channels=in_channels, 
        out_channels=out_channels,
        hidden_channels=hidden_channels,
        kernel_size=kernel_size,
        use_final_tanh=False,
        dilation=dilation)
    net.to(torch_device)

    link_mask, plaq_mask = next(masks)
    conditioner = csc.CSConditioner(net, n_knots=n_knots)
    plaq_coupling = csc.CouplingLayer(
        conditioner=conditioner,
        transform=csc.CSTransform(),
        mask=plaq_mask
    )
    link_coupling = U1GaugeEquivCouplingLayer(
        loops_function=loops_function,
        active_links_mask=link_mask,
        plaq_coupling=plaq_coupling)
    layers.append(link_coupling)

model = neumc.nf.flow_abc.TransformationSequence(layers)
\end{lstlisting}
We construct the layers one by one. For each layer, we create a neural network using an auxiliary function \verb|make_conv_net| passing in the previously defined parameters. We then create the conditioner passing in the network. In the current implementation, we use 2D convolutional networks, however, any other architecture is possible as far as the input and the output are of the correct size.  Next, we create the plaquette coupling layer. Apart from the conditioner and the circular splines transformation \verb|CSTransform|, we pass in the mask obtained from the \verb|masks| generator we have created at the beginning. 

We use this plaquette coupling layer to construct a gauge equivariant layer in the manner described in Section~\ref{sec:gauge-equivariant}. The constructor accepts additionally the link mask, which specifies which links are active and the loops function that will be used to calculate the additional invariants: the $\twobyone$ loops in this case.

And finally, we collect all the layers using the \verb|TransformationSequence| class. 

We still need the prior, which we define as uniform on the interval $[0,2\pi)$. 
\begin{lstlisting}
prior = neumc.nf.prior.MultivariateUniform(
    torch.zeros((2, *lattice_shape), device=torch_device),
    2 * torch.pi * 
        torch.ones(lattice_shape, device=torch_device)
)
\end{lstlisting}

Next, we set up the training
\begin{lstlisting}[]
from neumc.training.gradient_estimator \
    import REINFORCEEstimator

optimizer = torch.optim.Adam(
    model.parameters(), lr=0.00025)

batch_size = 512
n_batches  = 2

grad_estimator = REINFORCEEstimator(
    prior=prior,
    flow=model,
    action=qed_action,
    use_amp=False)
\end{lstlisting}
We choose the REINFORCE gradient estimator\cite{bialas2024} and \verb|Adam| optimizer. We set the batch size to 512 and the number of batches to two. This means that the gradient will be calculated on $2\times 512 =1024$ configurations, but this will be done in two batches of 512 configurations.  In this way, we can use a larger number of configurations to reduce the variance of the estimator and still fit on a GPU. The \verb|use_amp=False| parameter in the gradient estimator switches off the {\em automatic mixed precision}. Setting it to true would enable the use of the {\em tensor cores} on Nvidia cards. Those use half-float precision and may lead to numerical instability in some cases. 

After all this setup, the training loop is simple. The \verb|step| method of the gradient estimator does most of the work calculating the gradients, so we just have to call the \verb|step| method of the optimizer. 
\begin{lstlisting}
n_eras = 4
n_epochs_per_era = 50

for era in range(n_eras):
    total_ess = 0.0
    for epoch in range(n_epochs_per_era):
        optimizer.zero_grad()
        loss_, logq, logp = grad_estimator.step(
            batch_size=batch_size,
            n_batches=n_batches)
        total_ess += neumc.utils.ess(logp, logq)
        optimizer.step()
    total_ess /= n_epochs_per_era
    print(era, total_ess)    
\end{lstlisting}

We would like to stress that the implementation contains several components that can be varied independently within  reasonable constraints. First, we may vary the action. For example, we can obtain a pure gauge model by using the \texttt{U1GaugeAction} instead of \texttt{QEDAction}, but we may also use some form of improved action.  However, including a larger set of loops may require changing the masking patterns. 

Second, we may vary masking patterns and the loops used to calculate the invariants for the conditioner. However, we must be careful to match the masks with different loops.   

Third, we may change the transformation used in the flow. In this example, we are limited to non-compact projection and circular splines, but the users may implement their own transformations. 

Fourth,  we may use different gradient estimators, however, the choice depends on the transformation. If the transformation does not have the \texttt{reverse} method, we are limited to the \texttt{RTEstimator} (reparameterization trick).

And of course, we can change the conditioner, which would usually amount to changing the neural network architecture inside it.

\section{Summary and outlook}

In this paper, we presented \verb!neumc! a software package that allows users to implement 2D flow-based samplers for field theories. It is written using \verb!PyTorch! library. The package is constructed using separate modules for flow transformation, prior sampling, and gradient estimator, such that each of them can be modified separately without interfering with the others. At this point, the available physical models are: $\phi^4$ scalar theory and its vector version, XY model, and $U(1)$ gauge theory. We implemented affine coupling layers, non-compact projection layers, and circular splines. In the case of gauge theory, we constructed layers such that they are gauge equivariant, which is crucial for training efficiency. Finally, we provide the implementation for three gradient estimators that are used in the literature in the context of field theory samplers. 

There are several possible directions in which code can be developed further. First, sampling in more than two dimensions is required to cover more physically motivated models. Second, more complicated (in particular non-abelian) gauge groups should be implemented. Finally, a more efficient way for the treatment of fermions needs to be provided.
This is a subject of ongoing work. 

To conclude, we believe that the \verb!neumc! may be a useful tool for anyone interested in flow-based samplers.

\section*{Acknowledgements}

We thank Kim Nicoli and Christopher Anders for discussions. 
The computer time allocation 'plgnglft' on the Athena supercomputer hosted by AGH Cyfronet in Krak\'{o}w, Poland was used through the Polish PLGRID consortium. T.S. and D.Z. acknowledge the support of the Polish National Science Center (NCN) Grant No. 2021/43/D/ST2/03375. P.K. acknowledges the support of the Polish National Science Center (NCN) grant No. 2022/46/E/ST2/00346. D.Z. acknowledges the support of the Research Support Module under the program Excellence Initiative - Research University at the Jagiellonian University. This research was partially funded by the Priority Research Area Digiworld under the program Excellence Initiative – Research University at the Jagiellonian University in Kraków.

\appendix

\section{Installation}
\label{sec:installation}

Before using the software, the user must install the \verb!PyTorch! library. This step is heavily dependent on the operational system and hardware and is not described here. The software is distributed in the form of the git repository hosted on GitHub \cite{repo} that is composed of the \verb!neumc! python package and auxiliary files. After cloning the repository, the user can install the package using pip
\begin{verbatim}
cd <path-to-repo> 
pip install -e neumc
 \end{verbatim}
The optional \verb!-e! flag install the package in the developer mode, in this way any changes made to the package by the user will be immediately visible without reinstalling. The package requires additionally \verb!numpy! and \verb!scipy! packages to be installed. Now the user can start developing her own software based on the facilities provided by the \verb!neumc!. A good way to start are the scripts contained in the \verb!scripts! folder of the repository. 

A more pedagogical introduction is provided in the notebooks that can be found in the \verb|notebooks| folder. Notebooks are provided in the R markdown format (\verb|*.Rmd|) and have to be first converted to \verb|*.ipynb| files using e.g. \verb|jupytext| package. The notebooks require additionally the installation of the \verb|matplotlib.pyplot| package for plotting. 

\section{Masking patterns}
\label{sec:masking-patterns}

\subsection{Scalar masks}
\label{sec:scalar-masks}

As described in section \ref{sec:masking} the partition of configuration into the different parts needed by the coupling layers is done using masks. We provide several implementations of such masking patterns. 

\begin{figure}
    \centering
    \includegraphics[width=0.7\linewidth]{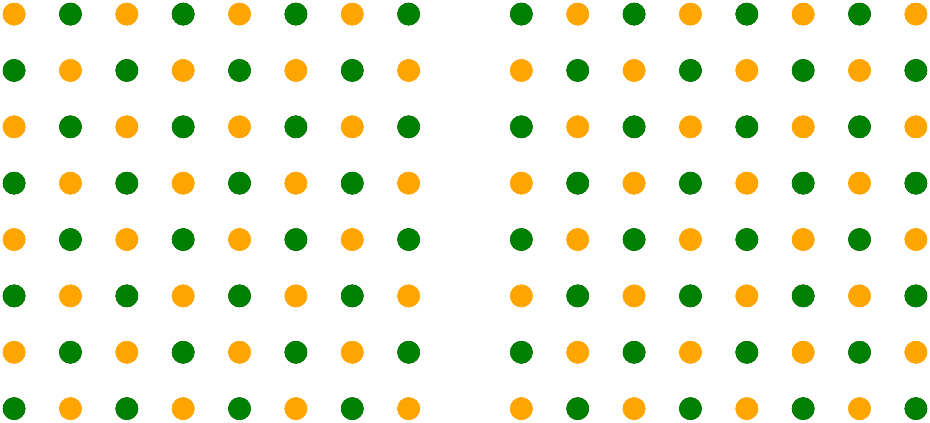}
    \caption{Checkerboard masks. Active sites are marked in orange and frozen sites in green. There are no passive sites. The period is two.}
    \label{fig:masks-checker}
\end{figure}

\begin{figure}
    \centering
    \includegraphics[width=0.7\linewidth]{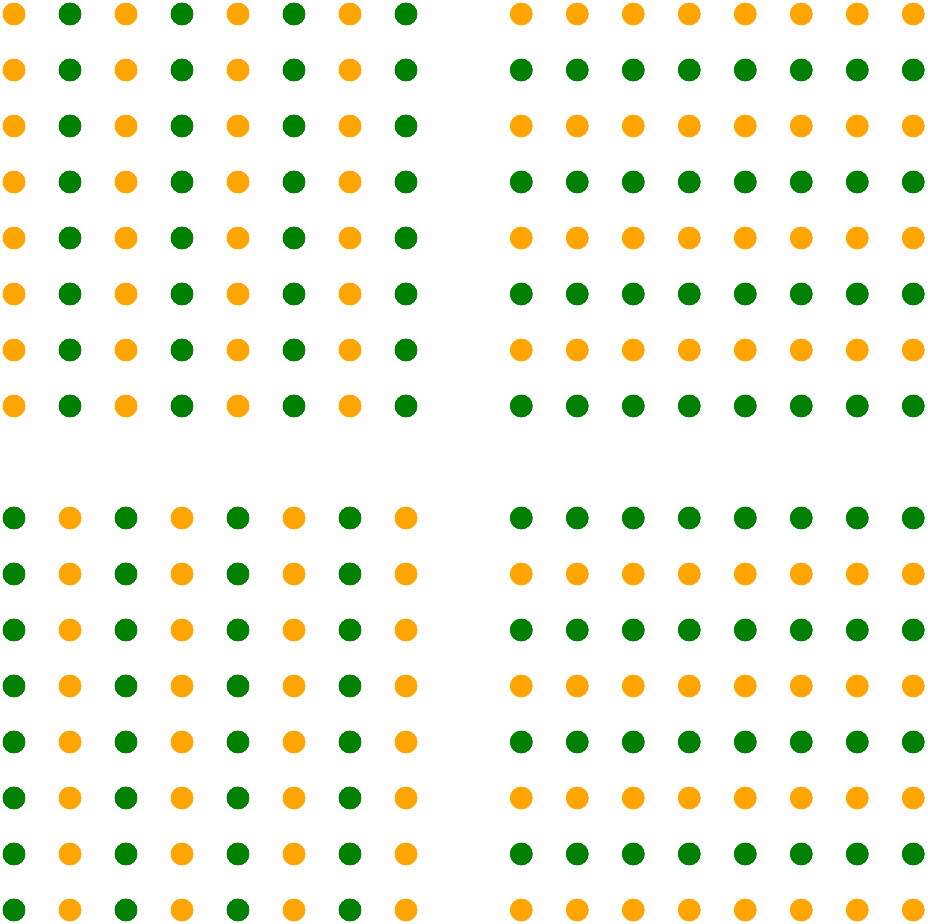}
    \caption{Stripes masks. Active sites are marked in orange and frozen sites in green. There are no passive sites. The period is four.}
    \label{fig:mask }
    \label{fig:masks-stripes}
\end{figure}

\begin{figure}
    \centering
    \includegraphics[width=0.7\linewidth]{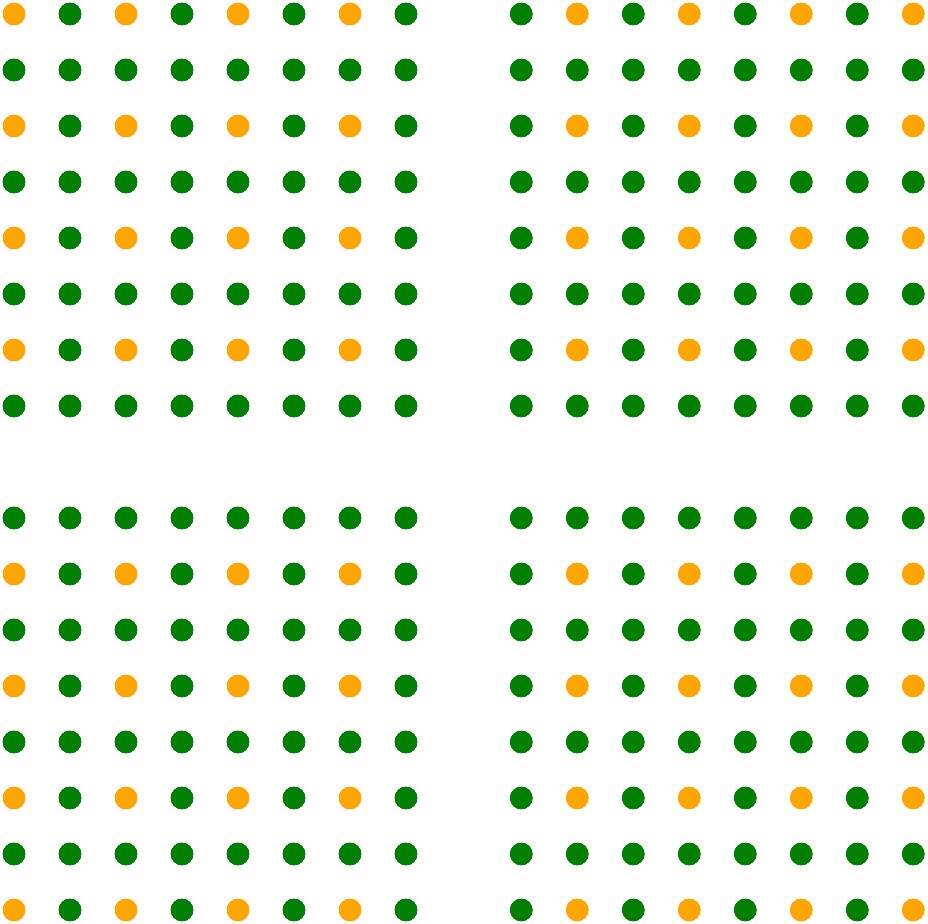}
    \caption{Tiled masks. Active sites are marked in orange and frozen sites in green. There are no passive sites. The period is four. Contrary to the previous masks the number of active and frozen sites is not equal.}
    \label{fig:masks-tiles}
\end{figure}

For scalar and vector fields we provide three different patterns of scalar masks. When using those masks on a vector field, all the components of the vector are masked in the same way. The patterns are: checkerboard, stripes and tiled masks that are presented in the Figures \ref{fig:masks-checker} to \ref{fig:masks-tiles}. 

\begin{figure}
    \centering
    \includegraphics[width=0.5\linewidth]{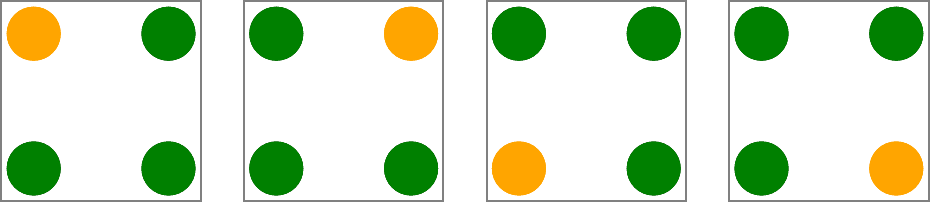}
    \caption{Tiles used to produce masks presented in Figure \ref{fig:masks-tiles}}
    \label{fig:simple-tiles}
\end{figure}

The tiled masks require a little bit of an explanation. Actually, it is a family of masks defined by a set of tiles. Each mask is constructed by taking one tile from the set, and creating the active mask by tiling it to fill the desired configuration size. The masks in the Figure~\ref{fig:masks-tiles} were produced by the set of tiles presented in the Figure~\ref{fig:simple-tiles}. Please note that for those masks the numbers of frozen and active sites are not equal. 

\subsection{Gauge fields masks}
\label{sec:gauge-fields-masks}

\begin{figure}
    \centering
    \includegraphics[width=\linewidth]{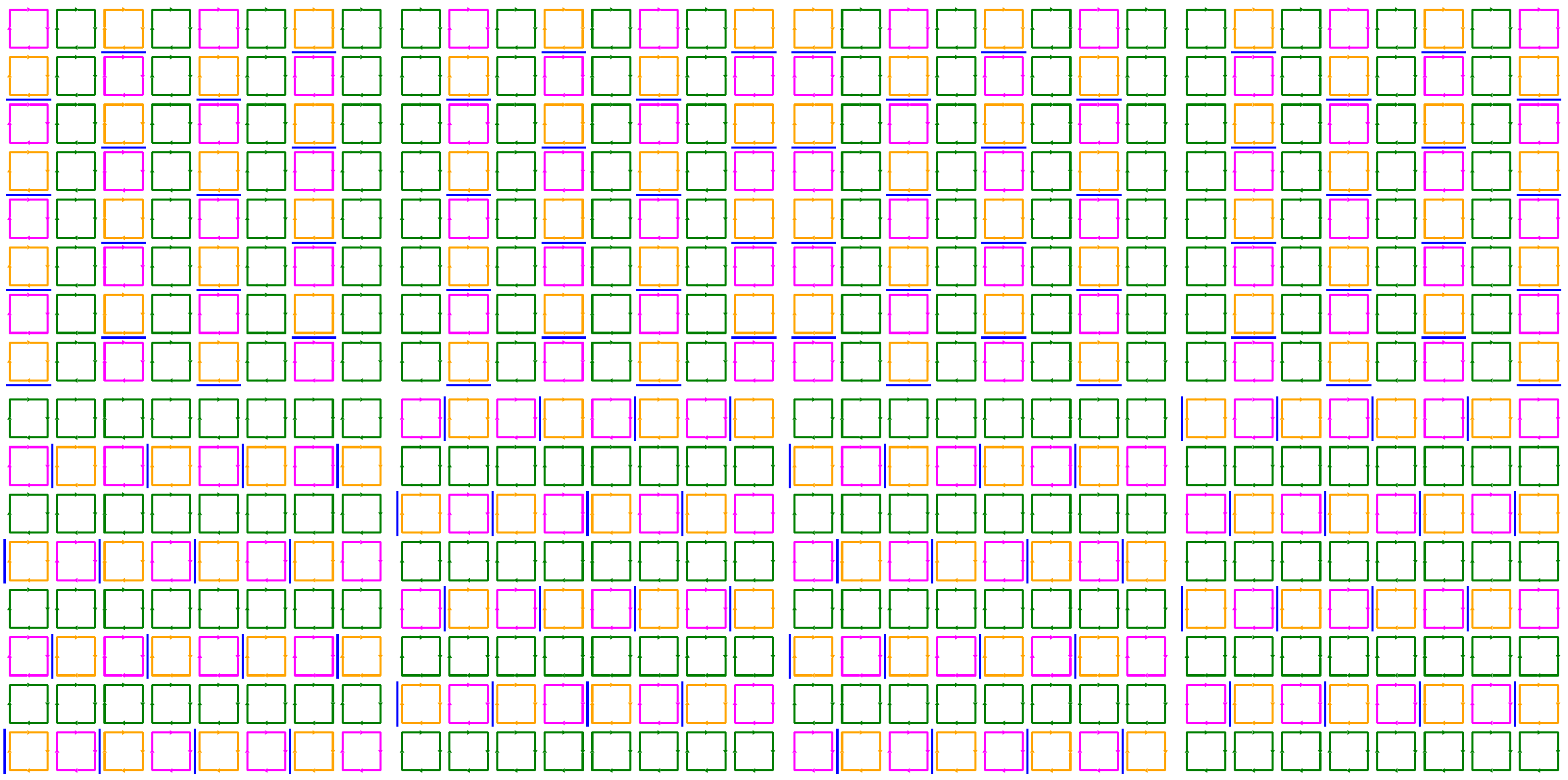}
    \caption{$U(1)$ masks. Active plaquettes are marked in orange, frozen sites in green, and passive in magenta. 
    Active links are marked in blue. 
    The period is eight.}
    \label{fig:masks-sch}
\end{figure}

As described in Section~\ref{sec:gauge-equivariant} masking patterns for gauge equivariant layers are more complicated. First of all, they involve two different sets of configurations: the links and the auxiliary loops. Consequently, we must provide two sets of masks: link masks and loop masks. Secondly, we may have different sets of loops, and we have to provide a mask for all of them. In Figure~\ref{fig:masks-sch} we present another masking pattern for the links and plaquettes more suitable for the Schwinger model \cite{Albergo2022}.  
If we want to extend this pattern to the $\twobyone$ loops we need additional masks. Those masks will have two channels, one for each orientation of loops. As explained earlier, we need only the frozen masks. Those will follow the frozen plaquettes masking. When the frozen plaquettes are organized in rows (second row in Figure \ref{fig:masks-sch}) we use this mask for the horizontal $\twobyone$ loops channel and zero for the other channel, and similarly for the masks where plaquettes are organized in columns (first row in Figure \ref{fig:masks-sch})

\section{Rational splines}
\label{sec:rational-splines}

First, it is important to note that we have a different spline for every different configuration $\bvphi$ and each lattice site in those configurations. So for each batch, the splines are defined by a set of three tensors $\bm{x}$, $\bm{y}$ and $\bm{s}$ of dimension $(N_b,L_x,L_y,N_k)$ each, where $N_k=K+1$ is the number of knots.

Evaluating the $h(\vphi)$ for arbitrary $\vphi\in[0,2\pi)$ requires finding such $k$ that $\u{x}k\leq\vphi<\u{x}{k+1}$. This is performed by a function \verb|torch.searchsorted|.  This function will take tensor $\bm{x}$ of size $(N_b,L_x,L_y,N_k)$ 
 and the batch of configurations $\bvphi$ "unsqueezed" to dimension  $(N_b,L_x,L_y,1)$. It would return a tensor $\bm{k}'$ that we "squeeze" back to the dimension  $(N_b,L_x,L_y)$ such that
 \begin{equation}\label{eq:indexing}
     x_{n,i,j,(k'_{n,i,j}-1)} \leq\vphi_{n,i,j}< x_{n,i,j,k'_{n,i,j}}
 \end{equation}
Given the tensor $\bm{k}=\bm{k}'-1$ we have to use it to index the tensor $\bm{x}$. This can be done using the so-called {\em smart indexing}. This works as follows: given a tensor $\bm{x}$ of dimension $(N_b,L_x, L_y,N_k)$ and four integer tensors $(\bm{i}_0, \bm{i}_1, \bm{i}_2, \bm{i}_3)$ of dimension $(N_b,L_x, L_y)$ we can create a new tensor
$\bm{x}[\bm{i}_0,\bm{i}_1,\bm{i}_2,\bm{i}_3]$,  of dimension $(N_b,L_x, L_y)$, such that
\begin{equation}
\begin{split}
    \bm{x}&[\bm{i}_0,\bm{i}_1,\bm{i}_2,\bm{i}_3][i,j,l] \\
    &=  \bm{x}[\bm{i}_0[i,j,l], \bm{i}_1[i,j,l], \bm{i}_2[i,j,l], \bm{i}_3[i,j,l] ]
\end{split}    
\end{equation}
Comparing with \eqref{eq:indexing} we see that we require $\bm{i}_3=\bm{k}$ and
\begin{equation}
    \bm{i}_0[i,j,l]=i,\; \bm{i}_1[i,j,l]=j,\; \bm{i}_2[i,j,l]=l 
\end{equation}
Those tensors can be created using the \verb|arange| and \verb|expand| functions
\begin{lstlisting}
def make_idx(*dims):
    idx=[]
    for i,d in enumerate(dims):
        v = torch.ones(len(dims)).to(dtype=torch.int64)
        v[i]=-1
        ix = torch.arange(d).view(*v).expand(*dims)
        idx.append(ix)
    return tuple(idx)    
\end{lstlisting}
We can then combine this index with tensor $\bm{k}$ to index other tensors
\begin{lstlisting}
idx = make_idx(N_b,L_x,L_y)
k = torch.searchsorted(...)
ki = (*idx, k-1)
xi = x[ki]
\end{lstlisting}

\bibliographystyle{elsarticle-num} 
\bibliography{nmcmc}

\begin{thebibliography}{10}
\expandafter\ifx\csname url\endcsname\relax
  \def\url#1{\texttt{#1}}\fi
\expandafter\ifx\csname urlprefix\endcsname\relax\def\urlprefix{URL }\fi
\expandafter\ifx\csname href\endcsname\relax
  \def\href#1#2{#2} \def\path#1{#1}\fi

\bibitem{metropolis}
N.~Metropolis, A.~W. Rosenbluth, M.~N. Rosenbluth, A.~H. Teller, E.~Teller,
  \href{https://doi.org/10.1063/1.1699114}{Equation of state calculations by
  fast computing machines}, The Journal of Chemical Physics 21~(6) (1953)
  1087--1092.
\newblock \href {http://arxiv.org/abs/https://doi.org/10.1063/1.1699114}
  {\path{arXiv:https://doi.org/10.1063/1.1699114}}, \href
  {https://doi.org/10.1063/1.1699114} {\path{doi:10.1063/1.1699114}}.
\newline\urlprefix\url{https://doi.org/10.1063/1.1699114}

\bibitem{Hastings}
W.~K. Hastings, {Monte Carlo sampling methods using Markov chains and their
  applications}, Biometrika 57~(1) (1970) 97--109.
\newblock \href {https://doi.org/10.1093/biomet/57.1.97}
  {\path{doi:10.1093/biomet/57.1.97}}.

\bibitem{VANPRL}
D.~Wu, L.~Wang, P.~Zhang, Solving statistical mechanics using variational
  autoregressive networks, Phys. Rev. Lett. 122 (2019) 080602.

\bibitem{PhysRevD.100.034515}
M.~S. Albergo, G.~Kanwar, P.~E. Shanahan, Flow-based generative models for
  markov chain monte carlo in lattice field theory, Phys. Rev. D 100 (2019)
  034515.

\bibitem{Kanwar2020}
G.~Kanwar, M.~S. Albergo, D.~Boyda, K.~Cranmer, D.~C. Hackett, S.~Racanière,
  D.~J. Rezende, P.~E. Shanahan,
  \href{https://arxiv.org/abs/2003.06413}{Equivariant flow-based sampling for
  lattice gauge theory}, Physical Review Letters 125 (9 2020).
\newline\urlprefix\url{https://arxiv.org/abs/2003.06413}

\bibitem{Abbott:2023thq}
R.~Abbott, et~al., {Normalizing flows for lattice gauge theory in arbitrary
  space-time dimension} (5 2023).
\newblock \href {http://arxiv.org/abs/2305.02402} {\path{arXiv:2305.02402}}.

\bibitem{Albergo:2021bna}
M.~S. Albergo, G.~Kanwar, S.~Racani\`ere, D.~J. Rezende, J.~M. Urban, D.~Boyda,
  K.~Cranmer, D.~C. Hackett, P.~E. Shanahan, {Flow-based sampling for fermionic
  lattice field theories}, Phys. Rev. D 104~(11) (2021) 114507.
\newblock \href {http://arxiv.org/abs/2106.05934} {\path{arXiv:2106.05934}},
  \href {https://doi.org/10.1103/PhysRevD.104.114507}
  {\path{doi:10.1103/PhysRevD.104.114507}}.

\bibitem{Albergo2022}
M.~S. Albergo, D.~Boyda, K.~Cranmer, D.~C. Hackett, G.~Kanwar, S.~Racanière,
  D.~J. Rezende, F.~Romero-López, P.~E. Shanahan, J.~M. Urban,
  \href{http://arxiv.org/abs/2202.11712}{Flow-based sampling in the lattice
  schwinger model at criticality} (2 2022).
\newblock \href {https://doi.org/10.48550/arxiv.2202.11712}
  {\path{doi:10.48550/arxiv.2202.11712}}.
\newline\urlprefix\url{http://arxiv.org/abs/2202.11712}

\bibitem{Abbott:2022zhs}
R.~Abbott, et~al., {Gauge-equivariant flow models for sampling in lattice field
  theories with pseudofermions}, Phys. Rev. D 106~(7) (2022) 074506.
\newblock \href {http://arxiv.org/abs/2207.08945} {\path{arXiv:2207.08945}},
  \href {https://doi.org/10.1103/PhysRevD.106.074506}
  {\path{doi:10.1103/PhysRevD.106.074506}}.

\bibitem{Abbott:2024kfc}
R.~Abbott, A.~Botev, D.~Boyda, D.~C. Hackett, G.~Kanwar, S.~Racani\`ere, D.~J.
  Rezende, F.~Romero-L\'opez, P.~E. Shanahan, J.~M. Urban, {Applications of
  flow models to the generation of correlated lattice QCD ensembles}, Phys.
  Rev. D 109~(9) (2024) 094514.
\newblock \href {http://arxiv.org/abs/2401.10874} {\path{arXiv:2401.10874}},
  \href {https://doi.org/10.1103/PhysRevD.109.094514}
  {\path{doi:10.1103/PhysRevD.109.094514}}.

\bibitem{abbott2024practicalapplicationsmachinelearnedflows}
R.~Abbott, M.~S. Albergo, D.~Boyda, D.~C. Hackett, G.~Kanwar, F.~Romero-López,
  P.~E. Shanahan, J.~M. Urban,
  \href{https://arxiv.org/abs/2404.11674}{Practical applications of
  machine-learned flows on gauge fields} (2024).
\newblock \href {http://arxiv.org/abs/2404.11674} {\path{arXiv:2404.11674}}.
\newline\urlprefix\url{https://arxiv.org/abs/2404.11674}

\bibitem{PhysRevE.101.023304}
K.~A. Nicoli, S.~Nakajima, N.~Strodthoff, W.~Samek, K.-R. M\"uller, P.~Kessel,
  Asymptotically unbiased estimation of physical observables with neural
  samplers, Phys. Rev. E 101 (2020) 023304.

\bibitem{ETOinLFT}
K.~A. Nicoli, et~al.,
  \href{http://dx.doi.org/10.1103/PhysRevLett.126.032001}{Estimation of
  thermodynamic observables in lattice field theories with deep generative
  models}, Physical Review Letters 126~(3) (Jan 2021).
\newblock \href {https://doi.org/10.1103/physrevlett.126.032001}
  {\path{doi:10.1103/physrevlett.126.032001}}.
\newline\urlprefix\url{http://dx.doi.org/10.1103/PhysRevLett.126.032001}

\bibitem{Bialas:2024gha}
P.~Bia\l{}as, P.~Korcyl, T.~Stebel, D.~Zapolski, {R\'enyi entanglement entropy
  of a spin chain with generative neural networks}, Phys. Rev. E 110~(4) (2024)
  044116.
\newblock \href {http://arxiv.org/abs/2406.06193} {\path{arXiv:2406.06193}},
  \href {https://doi.org/10.1103/PhysRevE.110.044116}
  {\path{doi:10.1103/PhysRevE.110.044116}}.

\bibitem{Bulgarelli:2024yrz}
A.~Bulgarelli, E.~Cellini, K.~Jansen, S.~K\"uhn, A.~Nada, S.~Nakajima, K.~A.
  Nicoli, M.~Panero, {Flow-based Sampling for Entanglement Entropy and the
  Machine Learning of Defects} (10 2024).
\newblock \href {http://arxiv.org/abs/2410.14466} {\path{arXiv:2410.14466}}.

\bibitem{Wang:2023exq}
L.~Wang, G.~Aarts, K.~Zhou, {Diffusion models as stochastic quantization in
  lattice field theory}, JHEP 05 (2024) 060.
\newblock \href {http://arxiv.org/abs/2309.17082} {\path{arXiv:2309.17082}},
  \href {https://doi.org/10.1007/JHEP05(2024)060}
  {\path{doi:10.1007/JHEP05(2024)060}}.

\bibitem{Zhu:2025pmw}
Q.~Zhu, G.~Aarts, W.~Wang, K.~Zhou, L.~Wang, {Physics-Conditioned Diffusion
  Models for Lattice Gauge Theory} (2 2025).
\newblock \href {http://arxiv.org/abs/2502.05504} {\path{arXiv:2502.05504}}.

\bibitem{Nagai:2025rok}
Y.~Nagai, H.~Ohno, A.~Tomiya, {CASK: A Gauge Covariant Transformer for Lattice
  Gauge Theory}, PoS LATTICE2024 (2025) 030.
\newblock \href {http://arxiv.org/abs/2501.16955} {\path{arXiv:2501.16955}},
  \href {https://doi.org/10.22323/1.466.0030} {\path{doi:10.22323/1.466.0030}}.

\bibitem{Caselle:2022acb}
M.~Caselle, E.~Cellini, A.~Nada, M.~Panero, {Stochastic normalizing flows as
  non-equilibrium transformations}, JHEP 07 (2022) 015.
\newblock \href {http://arxiv.org/abs/2201.08862} {\path{arXiv:2201.08862}},
  \href {https://doi.org/10.1007/JHEP07(2022)015}
  {\path{doi:10.1007/JHEP07(2022)015}}.

\bibitem{Bulgarelli:2024brv}
A.~Bulgarelli, E.~Cellini, A.~Nada, {Scaling of Stochastic Normalizing Flows in
  $\mathrm{SU}(3)$ lattice gauge theory} (11 2024).
\newblock \href {http://arxiv.org/abs/2412.00200} {\path{arXiv:2412.00200}}.

\bibitem{Gerdes:2024rjk}
M.~Gerdes, P.~de~Haan, R.~Bondesan, M.~C.~N. Cheng, {Continuous normalizing
  flows for lattice gauge theories} (10 2024).
\newblock \href {http://arxiv.org/abs/2410.13161} {\path{arXiv:2410.13161}}.

\bibitem{bialas2024}
P.~Białas, P.~Korcyl, T.~Stebel,
  \href{https://www.sciencedirect.com/science/article/pii/S0010465524000171}{Training
  normalizing flows with computationally intensive target probability
  distributions}, Computer Physics Communications 298 (2024) 109094.
\newblock \href {https://doi.org/https://doi.org/10.1016/j.cpc.2024.109094}
  {\path{doi:https://doi.org/10.1016/j.cpc.2024.109094}}.
\newline\urlprefix\url{https://www.sciencedirect.com/science/article/pii/S0010465524000171}

\bibitem{bialas2022gradientestimators}
P.~Bialas, P.~Korcyl, T.~Stebel, {Gradient Estimators for Normalizing Flows},
  Acta Phys. Polon. B 55~(3) (2024) 3--A2.
\newblock \href {http://arxiv.org/abs/2202.01314} {\path{arXiv:2202.01314}},
  \href {https://doi.org/10.5506/APhysPolB.55.3-A2}
  {\path{doi:10.5506/APhysPolB.55.3-A2}}.

\bibitem{albergo2021introduction}
M.~S. Albergo, et~al., Introduction to normalizing flows for lattice field
  theory (2021).
\newblock \href {http://arxiv.org/abs/2101.08176} {\path{arXiv:2101.08176}}.

\bibitem{Nicoli:2023rcd}
K.~A. Nicoli, C.~J. Anders, L.~Funcke, K.~Jansen, S.~Nakajima, P.~Kessel,
  {NeuLat: a toolbox for neural samplers in lattice field theories}, PoS
  LATTICE2023 (2024) 286.
\newblock \href {https://doi.org/10.22323/1.453.0286}
  {\path{doi:10.22323/1.453.0286}}.

\bibitem{repo}
P.~Bialas, P.~Korcyl, T.~Stebel, D.~Zapolski,
  \href{https://github.com/nmcmc/neumc.git}{Neural monte-carlo} (2024).
\newline\urlprefix\url{https://github.com/nmcmc/neumc.git}

\bibitem{KL}
S.~Kullback, R.~A. Leibler, {On Information and Sufficiency}, The Annals of
  Mathematical Statistics 22~(1) (1951) 79--86.
\newblock \href {https://doi.org/10.1214/aoms/1177729694}
  {\path{doi:10.1214/aoms/1177729694}}.

\bibitem{Liu}
J.~S. Liu, Metropolized independent sampling with comparisons to rejection
  sampling and importance sampling, Statistics and Computing 6~(2) (1996)
  113--119.

\bibitem{Bialas:2021bei}
P.~Bia\l{}as, P.~Korcyl, T.~Stebel, {Analysis of autocorrelation times in
  neural Markov chain Monte Carlo simulations}, Phys. Rev. E 107~(1) (2023)
  015303.
\newblock \href {http://arxiv.org/abs/2111.10189} {\path{arXiv:2111.10189}},
  \href {https://doi.org/10.1103/PhysRevE.107.015303}
  {\path{doi:10.1103/PhysRevE.107.015303}}.

\bibitem{Kong}
A.~Kong, A note on importance sampling using standarized weights, University of
  Chicago Technical Reports (1992).

\bibitem{WilsonLatticeGauge}
K.~G. Wilson,
  \href{https://link.aps.org/doi/10.1103/PhysRevD.10.2445}{Confinement of
  quarks}, Phys. Rev. D 10 (1974) 2445--2459.
\newblock \href {https://doi.org/10.1103/PhysRevD.10.2445}
  {\path{doi:10.1103/PhysRevD.10.2445}}.
\newline\urlprefix\url{https://link.aps.org/doi/10.1103/PhysRevD.10.2445}

\bibitem{dinh2017density}
L.~Dinh, J.~Sohl-Dickstein, S.~Bengio, Density estimation using real nvp
  (2017).
\newblock \href {http://arxiv.org/abs/1605.08803} {\path{arXiv:1605.08803}}.

\bibitem{9089305}
I.~Kobyzev, S.~Prince, M.~Brubaker, Normalizing flows: An introduction and
  review of current methods, IEEE Transactions on Pattern Analysis and Machine
  Intelligence (2020) 1--1.

\bibitem{Kobyzev2019}
I.~Kobyzev, S.~J. Prince, M.~A. Brubaker,
  \href{https://arxiv.org/abs/1908.09257v4}{Normalizing flows: An introduction
  and review of current methods}, IEEE Transactions on Pattern Analysis and
  Machine Intelligence 43 (2019) 3964--3979.
\newblock \href {https://doi.org/10.1109/tpami.2020.2992934}
  {\path{doi:10.1109/tpami.2020.2992934}}.
\newline\urlprefix\url{https://arxiv.org/abs/1908.09257v4}

\bibitem{Durkan2019}
C.~Durkan, A.~Bekasov, I.~Murray, G.~Papamakarios,
  \href{http://arxiv.org/abs/1906.04032}{Neural spline flows} (6 2019).
\newblock \href {https://doi.org/10.48550/arxiv.1906.04032}
  {\path{doi:10.48550/arxiv.1906.04032}}.
\newline\urlprefix\url{http://arxiv.org/abs/1906.04032}

\bibitem{Dinh2014NICENI}
L.~Dinh, D.~Krueger, Y.~Bengio,
  \href{https://api.semanticscholar.org/CorpusID:13995862}{Nice: Non-linear
  independent components estimation}, arXiv: Learning (2014).
\newline\urlprefix\url{https://api.semanticscholar.org/CorpusID:13995862}

\bibitem{Dinh2016DensityEU}
L.~Dinh, J.~N. Sohl-Dickstein, S.~Bengio,
  \href{https://api.semanticscholar.org/CorpusID:8768364}{Density estimation
  using real nvp}, ArXiv abs/1605.08803 (2016).
\newline\urlprefix\url{https://api.semanticscholar.org/CorpusID:8768364}

\bibitem{Rezende2020}
D.~J. Rezende, G.~Papamakarios, S.~Racanière, M.~S. Albergo, G.~Kanwar, P.~E.
  Shanahan, K.~Cranmer, \href{https://arxiv.org/abs/2002.02428}{Normalizing
  flows on tori and spheres}, 37th International Conference on Machine
  Learning, ICML 2020 PartF16814 (2020) 8039--8048.
\newblock \href {https://doi.org/10.48550/arxiv.2002.02428}
  {\path{doi:10.48550/arxiv.2002.02428}}.
\newline\urlprefix\url{https://arxiv.org/abs/2002.02428}

\bibitem{Boyda2021}
D.~Boyda, G.~Kanwar, S.~Racanière, D.~J. Rezende, M.~S. Albergo, K.~Cranmer,
  D.~C. Hackett, P.~E. Shanahan,
  \href{https://journals.aps.org/prd/abstract/10.1103/PhysRevD.103.074504}{Sampling
  using su (n) gauge equivariant flows}, Physical Review D 103 (2021) 074504.
\newblock \href {https://doi.org/10.1103/PHYSREVD.103.074504/FIGURES/19/MEDIUM}
  {\path{doi:10.1103/PHYSREVD.103.074504/FIGURES/19/MEDIUM}}.
\newline\urlprefix\url{https://journals.aps.org/prd/abstract/10.1103/PhysRevD.103.074504}

\bibitem{pathGradient}
L.~Vaitl, K.~A. Nicoli, S.~Nakajima, P.~Kessel, Gradients should stay on path:
  better estimators of the reverse- and forward kl divergence for normalizing
  flows, Machine Learning: Science and Technology 3~(4) (2022) 045006.
\newblock \href {https://doi.org/10.1088/2632-2153/ac9455}
  {\path{doi:10.1088/2632-2153/ac9455}}.

\bibitem{Rezende2016}
A.~Mnih, D.~J. Rezende, Variational inference for monte carlo objectives, in:
  Proceedings of the 33rd International Conference on International Conference
  on Machine Learning - Volume 48, ICML'16, JMLR.org, 2016, p. 2188–2196.

\bibitem{Hackett:2021idh}
D.~C. Hackett, C.-C. Hsieh, S.~Pontula, M.~S. Albergo, D.~Boyda, J.-W. Chen,
  K.-F. Chen, K.~Cranmer, G.~Kanwar, P.~E. Shanahan, {Flow-based sampling for
  multimodal and extended-mode distributions in lattice field theory} (7 2021).
\newblock \href {http://arxiv.org/abs/2107.00734} {\path{arXiv:2107.00734}}.

\end{thebibliography}
 

\end{document}